\documentclass{aastex631}

\usepackage{array,multirow,graphicx}
\usepackage{float}
\usepackage{gensymb}

\graphicspath{{./}}

\begin{document}

\title{Mass and Metal Flows in Isolated IllustrisTNG Halos}

\author{Jacob Morgan}
\affiliation{University of Alabama\\
Tuscaloosa, AL, 35405}

\author{Jeremy Bailin}
\affiliation{University of Alabama\\
Tuscaloosa, AL, 35405}



\begin{abstract}

The cicumgalactic medium (CGM) is a reservoir of metals and star-forming fuel. Most baryons in the universe are in the circumgalactic medium (CGM) or intergalactic medium (IGM). The baryon cycle-- how mass and metals reach the CGM from the inner regions of the galaxy and how gas from the CGM replenishes star-forming activity in the inner regions-- is an essential question in galaxy evolution. In this paper, we study the flow of mass and metals in a stacked sample of 2770 isolated halos from the IllustrisTNG cosmological hydrodynamic simulation. The mean gas flow as a function of radius and angle is similar across a large galactic mass range when accounting for different feedback modes. Although both star formation and black holes cause powerful outflows, the flows from star formation are more angularly restricted. Black hole feedback dominates massflow throughout the halo, while star-formation feedback mainly affects the inner region. When scaling by virial radius ($R_v$), large dynamical changes occur at $0.2R_v$ for most halos, suggesting a characteristic size for the inner galaxy. Despite radio mode feedback from black holes being the primary quenching mechanism in IllustrisTNG, a small population of high mass radio mode disks are able to form stars. 

\end{abstract}




\section{Introduction} \label{sec:intro}

In 1956, Lyman Spitzer suggested the existence of pockets of cool Na I and Ca II gas in pressure equilibrium with hot, $T \sim 10^6$~K gas far from the stellar populations of galaxies. What he called the ``galactic corona'' we now call the circumgalactic medium (CGM) \citep{spitzer}. The CGM refers to the region of diffuse gas surrounding the bright, star-filled center of galaxies. Since the 1950s, the importance of this medium to galaxy evolution and as a reservoir for the Universe's metals and baryons has become clear, with the gas in virialized regions of galaxies having a density parameter of $\Omega \sim 0.024$ \citep{Fukugita_2004}. Since 2009, Hubble's Cosmic Origins Spectrograph (COS) has  allowed astronomers to study the CGM in much greater detail, and recent advancements in computing have allowed for powerful hydrodynamic simulations of the gas. Despite the considerable insight these tools offer, a full description of the CGM is still lacking; its position as the intermediary between the galaxy and the rest of the Universe means that a wide variety of processes affect, and are affected by, its nature. 

In particular we are interested in the flow of mass through the central galaxy and CGM. Many studies on Illustris, the Next Generation (hereafter TNG) have found the radio mode kinetic feedback of active galactic nuclei (AGN) is much more influential on star formation and gas dynamics than the thermal mode, even when the same amount of energy is released \citep{kineticFB1, kineticFB2}. \citet{ejective_preventative} analyzed the effects of AGN feedback on gas in TNG100 and TNG300 and found that the kinetic feedback not only heats the gas in the star-forming regions, but disperses it to higher radii, causing a large increase in entropy and cooling times. If the AGN feedback is still relatively weak, less entropy is deposited in the inner CGM-- this gas can then cool and fall to the central region, and the galaxy may not be quenched. However, a highly entropic inner CGM prevents this flow. Similarly, observations from the MANGA survey show that quenching is most closely associated with absolute velocity dispersion \citep{veldisp_quench}.

Although the flow of gas caused by feedback processes is dynamic, feedback may be important in determining the size of a galaxy. \citet{obs_gal_size} isolated a starburst galaxy in FIRE-2 and re-ran the simulation with different forms of AGN feedback beginning at the onset of rapid star formation. Without AGN feedback, both the stellar and gaseous halfmass radii shrink significantly over the course of only $35$~Myrs. Observables such as halflight radii are also affected, with longer wavelengths generally becoming more concentrated, and dust obscuring higher frequency light. The addition of AGN feedback can maintain or even increase the stellar and gaseous halfmass radii depending on the chosen model, and less star formation occurs, leading to less dust. 

 To gain insight into the impact galactic properties have on galactic and CGM gas we use a sample of 2770 isolated halos from Illustris TNG100-1 and calculate their average instantaneous gas dynamics as a function of radius and polar angle. Section 2 covers Illustris TNG100 and its properties. In Section 3 we introduce our halo selection and analysis methods. We present our results in Section 4, and in Section 5 we discuss our conclusions and future work.

\section{Illustris TNG} \label{sec:TNG}
In this section we cover the simulation used for our work in more detail. IllustrisTNG uses the AREPO algorithm, which combines aspects of smooth particle hydrodynamics (SPH) and adaptive mesh refinement (AMR) codes \citep{gadget2}. In the AREPO code, points float through the simulation naturally following density, similar to SPH. The positions of these points are recorded and multifaceted borders are drawn between each, creating an amorphous grid of gas cells. Mass and metals can flow through the simulation both from the motion of cells, and also via transfer across cell boundaries. The points move and the cells are redrawn each timestep. This avoids sudden changes in resolution (as are present in AMR codes) and the use of a smoothing kernel (as in SPH codes) \citep{arepo}.

IllustrisTNG consists of three different volumes and several resolution levels for each volume \citep{unbind, BH_quench, TNG_paper3, TNG_paper4, TNG_paper5}. TNG50 has the highest resolution, but smallest volume and is thus best suited for studying small populations of galaxies in detail. TNG300 is lower resolution but contains much more massive objects and can be used for studying large scale structure. For this work, we used TNG100-1, which strikes a balance between these and allows us a larger sample size of galaxies. The average gas cell size and the gravitational softening length at $z=0$ are $15.8$ and $0.18$~kpc, respectively. 

The initial conditions of the dark matter and gas particles are meant to reproduce the Planck 2016 results at $z=0$ \citep{tng_public, Planck2016}. Each simulation records 100 snapshots from $z \approx 20$ to $z \approx 0$. For each snapshot, halos are found using a friend-of-friends (FoF) algorithm. The subhalos within them are identified using the ``Subfind'' algorithm, based on a gravitational unbinding procedure. Particles bound to the object at the center of the FoF halo form the central subhalo, and those bound to other points form satellite subhalos. Particles that are placed in groups by the FoF algorithm but not the subfind algorithm (i.e., within a halo but not bound) are called ``inner fuzz'' \citep{unbind}. There is also ``outer fuzz'', which are particles near the halo, even within the virial radius, but not assigned to it via the FoF algorithm. The outer fuzz is often not included in Illustris papers, but we do include it here. Preliminary results show that particles of this type are spatially well-mixed with particles belonging to the halo, and contribute significantly to the rate of outflowing material past $0.8R_v$ \citep{MINE2}. 


The physical model in TNG includes the following properties \citep{tng_public}: 1) A uniform ionizing radiation field  that changes with redshift, 2) Radiative cooling, including from metal lines, 3) Self-shielding 4) Star formation, which begins stochastically in gas cells above a certain density threshold, 5) star-forming gas cells governed by an effective equation of state that determines the amount of pressure they exert on the ISM, 6) Chemical enrichment of star particles and nearby gas with time, accounting for winds and mass loss from core collapse and white dwarf supernovae, and neutron star mergers, 7) Stellar feedback with energy-driven winds, 8) Black holes; when a halo reaches a mass of $8 \times 10^{10}\mathrm{M_{\odot}}/h$ it is seeded with a black hole of mass $8 \times 10^5 \mathrm{M_{\odot}}/h$ \citep{tng_gal_form} 9) Active galactic nuclei (AGN) feedback in two modes; ``radio/kinetic mode'' at low accretion rates and ``quasar mode'' at high accretion rates. In the low accretion mode, the AGN gives mechanical feedback in the form of outward pressure to nearby cells, while in the quasar mode it directly injects thermal energy to the cells \citep{major_merger}. 10) Primordial magnetic fields around overdensities at $z=127$, which are amplified by star formation and grow with time to galactic magnetic fields. 

Like most simulations, the subgrid models that govern behavior too detailed to simulate in TNG are ``tuned'' to reproduce certain observables at $z=0$; the stellar-mass-to-stellar-halfmass-radius relation, black-hole-mass-to-stellar-mass relation, gas mass fraction within the virial radius of massive halos, stellar-to-halo mass relation, stellar mass function, and the star formation rate density from $0<z<10$. IllustrisTNG also reproduces many observables independent of its tuning, including quenched galaxies at low and high redshift, the optical morphologies of galaxies, and the existence of uncommon galaxies such as those with low surface brightness or stripped of gas via ram pressure. TNG100 and TNG300 have been shown to produce OVI in the CGM that fits observations. Although most simulations correctly predict the lower covering fraction of high column density lines of sight in passive galaxies, they generally underpredict the amount of OVI around galaxies of both types \citep{TNG_OVI}. Oxygen, and especially OVI, is an indicator of star formation, with COS finding few passive galaxies with $N_{OVI}>10^{14} \mathrm{cm^{-2}}$ \citep{O6}. In the ENZO simulation, \citet{cen} finds that OVI is a short-lived state in the CGM, requiring consistent energy injection from the central galaxy to maintain. They find that active galaxies can have up to 30\% of their oxygen in this state, while passive galaxies have only 5\%. The oxygen content of TNG galaxies is similar to that found COS observations. TNG100 and TNG300 correctly predict the bimodal distribution of OVI, with star-forming galaxies having significantly more in their CGM \citep{TNG_OVI}. However, despite good agreement on covering fractions between ENZO and observations, \citet{foggie} find that the actual structure of the CGM within ENZO can be highly resolution dependent, and that similar covering fraction profiles can disguise differences in cloud size, number, and composition.

A study of AGN in TNG300 shows good agreement between the simulation and observations at low redshift. \citet{TNG_AGN} find that at low black hole mass, the primary growth channel for AGN is accretion (i.e., they are generally in the high accretion, quasar feedback mode) until the feedback becomes strong enough to effectively prevent the cooling/infall of gas. The AGN then enters the low accretion kinetic radio mode, hereafter ``RM'', and the feedback pressure becomes even stronger, preventing not only further black hole accretion, but star formation. Conversely, the quasar mode feedback has almost no effect on star formation rate. Because the AGN ``low accretion'' and ``high accretion'' modes are determined by the percent black hole mass accreted, it is significantly harder for more massive black holes to reach the high accretion state. AGN transition from quasar mode to radio mode at a mass of $M_{AGN}\sim 10^8$. Once the radio power reaches $>10^{42.5}\mathrm{erg s^{-1}}$ the halo is almost guaranteed to be quenched. The importance of this feedback is reflected in a significant narrowing of the stellar mass-black hole mass relationship after $M_{AGN}\sim 10^8$; because the quenching is associated with a lack of AGN accretion, both the AGN and stellar mass can now grow only through mergers. This process is the primary quenching mechanism in TNG. Combined with the stellar-black-hole mass relation, this makes both AGN feedback and quenching in TNG highly mass dependent.

The effects of the AGN RM feedback may also be seen in the morphology-color relation in TNG. When compared with the Pan-STARRS dataset, we see the well known, ``blue disk, red ellipticals'' paradigm is not reproduced well in the simulation. Instead, color is strongly linked only to stellar mass, showing little connection to several morphological measures. We note that the original TNG simulation-- which did not have the RM feedback at low accretion rates-- also did not have this red spiral problem, producing a morphology-color relation similar to that observed in Pan-STARRS \citep{red_spirals}. This supports the idea that the issue may be due to the efficient quenching of spirals once their AGN is massive enough to drop into the low accretion mode.

\section{Methods} \label{sec:methods}
\subsection{Halo Selection Methods}
Our initial sample consists of isolated halos at redshift $z \approx 0.2$. Isolated halos are found by analyzing changes in the primary subhalo gas mass and stellar mass within $2R_{1/2}$, where $R_{1/2}$ is the stellar halfmass radius. If central stellar mass or total gas mass changes by $10 \%$ or more for 5Gyr before $z=0$, the halo is not considered isolated. This accounts for very gas-rich mergers. Total gas mass is used instead of central gas mass because nearly all primary subhalos experience large changes ($>30 \%$) in central gas mass over 5Gyr. This agrees with our ultimate results, which indicate that without infalling gas the inner region of many halos could be evacuated within a Gyr. We then removed several ``fly-by'' events, which occur when two halos interact but their central subhalos do not coalesce, so no merger is recorded in the sublink tree. This leaves us with a sample of 2770 halos. 

We divide these halos by total mass into brackets from $11<\log(M/M_{\odot})<12.5$ in increments of $0.5$. We then split all halos in each mass bracket into various categories: disk/spheroid, star-forming/quiescent, and strong/weak RM AGN feedback, and cross-categories. This allows us to isolate the effects of different types of feedback. We will analyze the gas and metal dynamics of quiescent and RM-quiet ``low feedback'' halos (LFB), star-forming and RM-quiet halos (SF), quiescent and RM-loud halos (RM), and halos with both star formation and RM-loud AGN (SFRM), and disks (D) and spheroids (S). Although the kinetic radio mode is nominally the ``low activity'' state for AGN, it has been found this mode is actually more efficient at moving energy through the gas \citep{ejective_preventative, kineticFB2, xray_metal_line_emission}. We determine the morphology of the sample using the fraction of stars in the central subhalo which have a circularity fraction above 0.7 (`CircAbove07Frac', a pre-calculated measure in TNG for subhalos with $M_*>3.4 \times 10^8$).

Although halo properties are continuous, bimodal distributions are common. For each category (morphology, star formation, and RM activity) we choose a single threshold value that meaningfully subdivides the sample while maintaining sufficient size in each mass bracket. For morphology, we chose a threshold of $CircAbove07Frac>0.3$, i.e., a halo is categorized as ``disk'' when $30\%$ or more of its stars have a circularity above $0.7$.

Star formation rate is measured many different ways in TNG; star formation within $2R_{1/2}$ and total SFR, as well as being measured across different timescales. For each metric we used the same threshold ($10^{-10}/\mathrm{yr}$) for specific star formation rate (sSFR). We tested our results using different timescales for both SF and RM feedback and found the feedback activity averaged over the past $200$~Myrs to be the most predictive in terms of bulk gas flow. We will explore differences between halos with recent star formation and RM activity ($200$~Myrs) vs more distant activity ($1$~Gyr ago) in a subsequent paper.

\begin{figure*}[h]
\hspace{-0.25cm}
\includegraphics[width=1.0\textwidth]{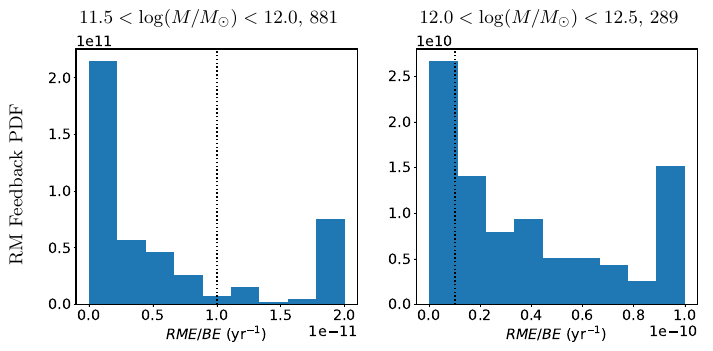}
\caption{Histograms of the ratio of RM AGN activity to binding energy. Left: RM activity of halos with $11.5<\log(M/M_{\odot})<12.0$. Right: RM activity of halos with $12.0<\log(M/M_{\odot})<12.5$. Only halos with non-zero radio mode energy are included. Our RM-active threshold of $10^{-11}$~/yr is denoted by the vertical dotted line. A long tail in the high mass distribution has been truncated and all placed in the highest bin.}
\label{fig:RMAGN}
\end{figure*}

Our star formation rates and morphology are based on pre-calculated figures in TNG. However, only total cumulative energy production is recorded for quasar and radio mode AGN. We use the cumulative energy given off by the central black hole between the target snapshot and the one immediately prior and divide by the exact time separating snapshots ($\sim 184$~Myrs) to get the average energy emitted in a year, $RME$. We also calculate the total binding energy ($BE$) of each halo. A halo is considered ``RM'' if $RME/BE>10^{-11}$~/yr. We use the binding energy ratio as we are particularly interested in the AGN's ability to quench halos, i.e., evacuate the central region, strip them of gas, or disperse and heat the gas. Almost no halos in the lowest mass bracket ($11<\log(M/M_{\odot})<11.5$, 1/1600 halos) qualify as an RM halo and very few in the middle bracket ($11.5<\log(M/M_{\odot})<12$, approximately $5\%$ of 881 halos) qualify, but over half of halos in the most massive group ($12<\log(M/M_{\odot})<12.5$) have strong RM activity. Figure \ref{fig:RMAGN} shows the average radio mode power released during the last $\approx 200$~Myr. Even though our threshold for RM activity increases with mass, more massive halos are more likely to have strong RM AGN. 

\subsection{Analysis Methods} \label{subsec:analysis}
We take a volumetric approach to the data; all gas and wind particles within the virial radius are collected for each halo. This bypasses the FoF algorithm IllustrisTNG uses to define halos, and allows us to include the outer fuzz particles. These particles are often not included in studies of TNG halos, but we find that they have a prominent effect on gas dynamics near the virial radius, especially for $\log(M/M_{\odot}) \approx 12-12.5$ halos. A forthcoming paper will explore the effects of the outer fuzz specifically.  Wind particles are classified as a variety of star particle in IllustrisTNG, but are included in our analysis as they transport mass and metal. Henceforth, unless otherwise noted, ``gas'' refers to both gas and wind particles.

After separating the halos into types as described in the previous section, we combine halos of the same mass/type. To do so we first create a common coordinate system for all halos. We align the z-axis with the angular momentum vector of the stars and gas of the central subhalo within $2R_{1/2}$ and normalize the radial coordinate by the virial radius, $R_v$. We then measure the angle from the z-axis, with $\theta=0\degree$ corresponding to the angular momentum minor axis and $\theta=90\degree$ corresponding to the major axis. After binning the data by normalized radius, polar angle, or both for 2D plots, the relevant properties in each bin are calculated and stacked by halo mass and feedback mode. Because our data is gathered from many halos the results are population averages and may not be applicable to individual halos.

\begin{figure*}[h]
\includegraphics[width=1.0\textwidth]{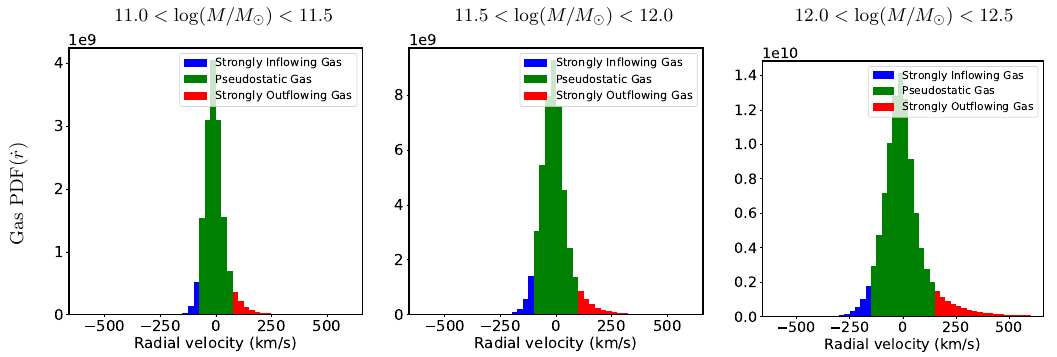}
\caption{Velocity distribution of the gas mass.}
\label{fig:inflow_outflow_histograms}
\end{figure*}

 To facilitate a deeper understanding of the mass and metal flows, we separate the particles into different dynamic types based on radial velocity; ``strongly inflowing'', ``pseudostatic'', and ``strongly outflowing''. We use a radial velocity threshold of ($75$km/s, $100$km/s, $150$km/s) for the mass bracket ($11.0<\log(M/M_{\odot})<11.5$, $11.5<\log(M/M_{\odot})<12.0$, $12.0<\log(M/M_{\odot})<12.5$) to define strongly inflowing/outflowing. Figure \ref{fig:inflow_outflow_histograms} shows velocity histograms for our sample. Each mass bracket has a roughly equal fraction of pseudostatic gas, and the long tail of outflowing gas in the rightmost plot hints at the strong, mass-dependent radio mode feedback common in the highest mass range. For each bin we calculate properties of interest for each particle type, i.e., metallicity of the inflowing, pseudostatic, and outflowing gasses. 

We are interested in the bulk flow of gas between the CGM and central galaxy. The rate of mass flow ($MF$) across each bin is calculated for each halo like so: 

$$MF_X=\frac{\Sigma_i^N m_{X, i}{\dot{r}}_i}{\Delta r}$$
where $N$ is the number of particles of dynamic type $X$, $m_{X, i}$ is the mass and $\dot{r_i}$ the radial velocity of each particle of dynamic type $X$ in the bin, and $\Delta r$ is the width of the radial bins for this halo. A positive (negative) value indicates the mass in this bin is outflowing (inflowing) on average. This value is then simply averaged for all halos of a particular mass and type. This method is repeated with metal mass in each bin to find the metalflow. The difference in massflow between each bin is also calculated, and represents the net accumulation of mass, or ${\dot{M}}_X$; a positive (negative) value indicates the mass within that bin is increasing (decreasing). Mass accumulation is proportional to the negative radial derivative of the massflow of gas and wind particles; if the flow is more outward at larger radius, the region is being evacuated, while if the flow is more inward at large radius the region is experiencing an increase in mass. Mass and metallicity accumulation calculated this way does not account for the effects of star formation or other processes which may generate or consume particles. 

The rate of change in the metallicity due to gas of type $X$ can be found from the mass and metallicity accumulation rates of that type:

$$\left(\frac{\mathrm{d}}{\mathrm{d}t}\log(Z/Z_{\odot}) \right)_X=\frac{1}{\ln10}\left(\frac{{\dot{M}}_{Z,X}}{M_Z}-\frac{{\dot{M}}_X}{M}\right)$$
where $M_X$ and $M_{Z,X}$ represent the mass and metalmass of dynamic type $X$ in the bin and $M$ and $M_z$ represent the total mass and metalmass in the bin. Finally, in a spherical 3D mapping, bin volume varies with polar angle:

$$V_{bin} = 2\pi \int_{\theta_0}^{\theta_1} \int_{r_0}^{r_1} r^2 \sin\theta \,dr \,d\theta=\frac{2\pi}{3}(r_1^3-r_0^3)(\cos\theta_1-\cos\theta_0)$$
To create isotropic equivalents we adjust the values in each bin by the ratio of its volume to the volume of a spherical shell at the same radius, $F=V_{bin}/V_{shell}=(cos\theta_1-cos\theta_0)/2$, and $P_{iso}=P/F$, where $P$ is any mass, massflow, or $\dot{M}$ property.

\section{Results} \label{sec:results}
\subsection{Results: All Galaxies} \label{subsec:results_all}
\begin{figure*}[h]
\includegraphics[width=1.0\textwidth]{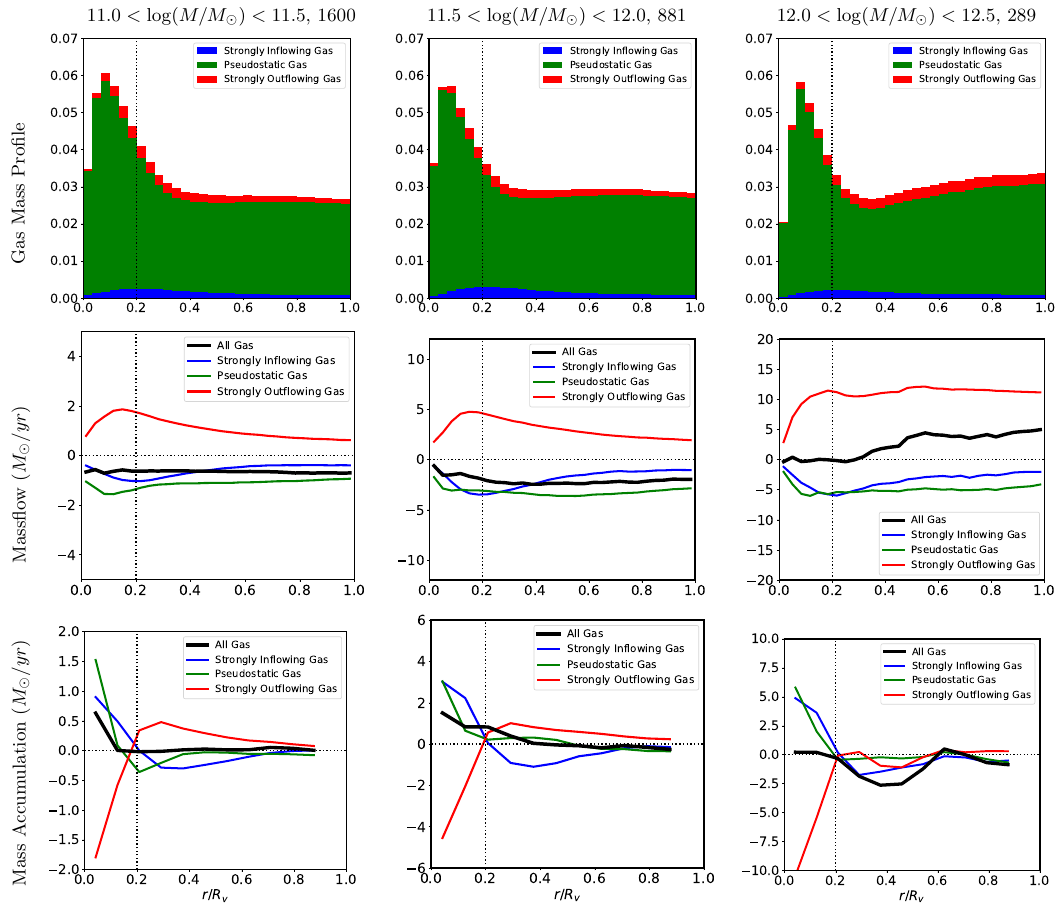}
\caption{Top: PDFs of the three different gas types as a function of radius, normalized by virial radius. Because the volume of each bin is proportional to $r^2$, a flat distribution corresponds to a $r^{-2}$ density profile. Middle: The radial flow of mass caused by each type of gas. Bottom: Mass accumulation as a function of normalized radius. The number at the top right of each column indicates how many halos are in that mass range.}
\label{fig:all_massfrac_massflow_mdot}
\end{figure*}

There are many factors to consider when describing the behavior of gas in TNG halos. As a baseline we will begin by looking at the mass dependence. Figure \ref{fig:all_massfrac_massflow_mdot} shows the radial distribution (top), flow of gas (middle) and mass accumulation (bottom) for our entire sample. The dotted vertical lines denote $0.2R_v$. In the top row, we see that the vast majority of all gas is pseudostatic, but that the most massive halos have more outflowing gas, especially at high radius. Meanwhile, they also have less inflowing gas throughout. Halos in the lower two mass brackets have a relatively flat distribution past $0.4R_v$, implying the density falls off as $\sim 1/r^2$. But the largest halos have a positive slope in this radial range, so the density decreases more slowly.

In the middle row of the Figure, we see that flow rate both in and out increases with mass. The strongly inflowing gas is negative at all radii by definition, but we also see the pseudostatic gas is net inflowing at all radii for all masses. This is true for every subsample; pseudostatic gas is always net inflowing on average. The peak inflow and outflow rates occur near $\sim 0.2R_v$. When looking at the net behavior the gas, the lowest and middle mass halos nearly always have net inflows (see Sections \ref{subsec:results_RM} and \ref{subsec:results_SFRM} for exceptions) with the flow rate becoming more extreme due to the deeper gravitational well. The total gas behavior closely resembles the pseudostatic gas behavior. However, despite the most massive halos having similar radial profiles for inflowing and pseudostatic gas, the extreme outflows decouple the total gas dynamics from the pseudostatic gas dynamics and there is a net outflow everywhere past $0.3R_v$.

In the bottom row of the figure we plot the accumulation of mass; differences in massflow lead to mass building up in certain bins, proportional to the negative of the massflow slope. However, it should be noted that particles may change dynamic type (for instance ``strongly outflowing'' to ``pseudostatic'', as is the case for the lowest mass bracket near $0.2R_V$), causing more gas of that type to enter a region than leave, and giving the appearance of mass accumulation; only the total net accumulation is entirely physical. At all halo masses we see that the innermost region of the galaxy is rapidly stripped of mass by high velocity outflows and rapidly refilled by a combination of high velocity inflows and pseudostatic gas. Inflowing gas begins to slow significantly near $0.2R_v$, and outflowing gas originates from within this radius, giving an approximate size of the central subhalo, $R_c=0.2R_v$. For the lowest and middle mass bins, the total behavior of the gas (black) is to inflow at all radii and accumulate at low radii. This is consistent with simulations showing halos in this mass range are still experiencing gas-rich mergers \citep{Hopkins_2009}. Although the inflows are stronger at low radii for the highest mass bracket, the outflows prevent significant accumulation of mass. At all masses there is a slight accumulation of outflowing gas only at $\sim 0.2R_v$, which corresponds to the peak in massflow. Past this radius there is some outflowing mass accumulated as the particles interact with the CGM and slow, giving the ``strongly outflowing'' massflow and mass accumulation plots both a negative slope. However, the strong outflow rate in the highest mass bracket actually increases slightly with radius, and we see no net mass accumulation-- instead, these halos are losing mass at nearly all radii past $0.2R_v$, indicating gas is expelled entirely from the halo.

\begin{figure*}
\includegraphics[width=1.0\textwidth]{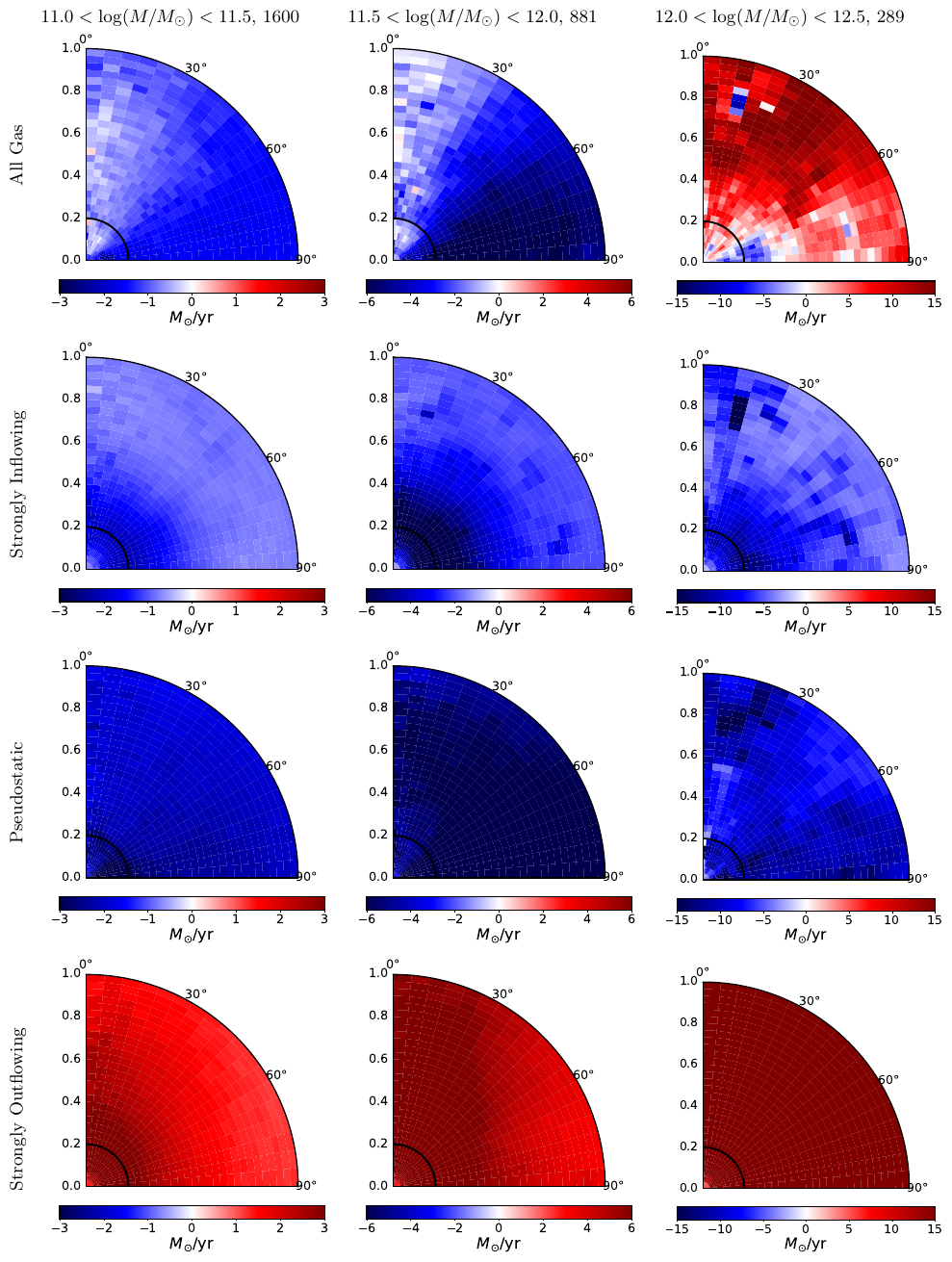}
\caption{Gas flow intensity as a function of normalized radius and polar angle. Blue is inflowing, red outflowing. Left to right: the lowest to highest mass brackets. Top to bottom: The flow of all gas, strongly inflowing gas, pseudostatic gas, and strongly outflowing gas.}
\label{fig:all_2d_massflow}
\end{figure*}

Figure \ref{fig:all_2d_massflow} shows the two dimensional distribution of gas flows. Looking at the total flow rates in the top row, the outflows are more intense near the poles, which fits observations of both star-forming \citep{SF_bipolar_10kpc} and radio loud galaxies \citep{RM_bipolar}. The lower two mass brackets have a clear bipolar structure, with neutral flows along the poles and inflows elsewhere. Between the lowest and middle mass brackets we see an intensification of the inflowing behavior everywhere besides the poles. This situation is reversed for the highest mass bracket, which has outflows everywhere except far from the polar regions. Despite these differences in overall gas flow, when we break the gas down into its constituent parts we see the strongly inflowing, strongly outflowing, and pseudostatic gasses are quite similar across mass: the inflowing gas is strongest within $0.4R_v$ for the lower two mass brackets, but becomes less intense at the very innermost region, where pseudostatic gas is more dominant; the flow of pseudostatic gas is essentially constant with radius, though the flow is disrupted in the innermost region of high mass halos; the outflowing gas is bipolar even in LFB halos, and is ostensibly responsible for the disruption. The rate of strongly outflowing gas also remains high out to a larger radius than strongly inflowing gas. 

\begin{figure*}[h]
\includegraphics[width=1.0\textwidth]{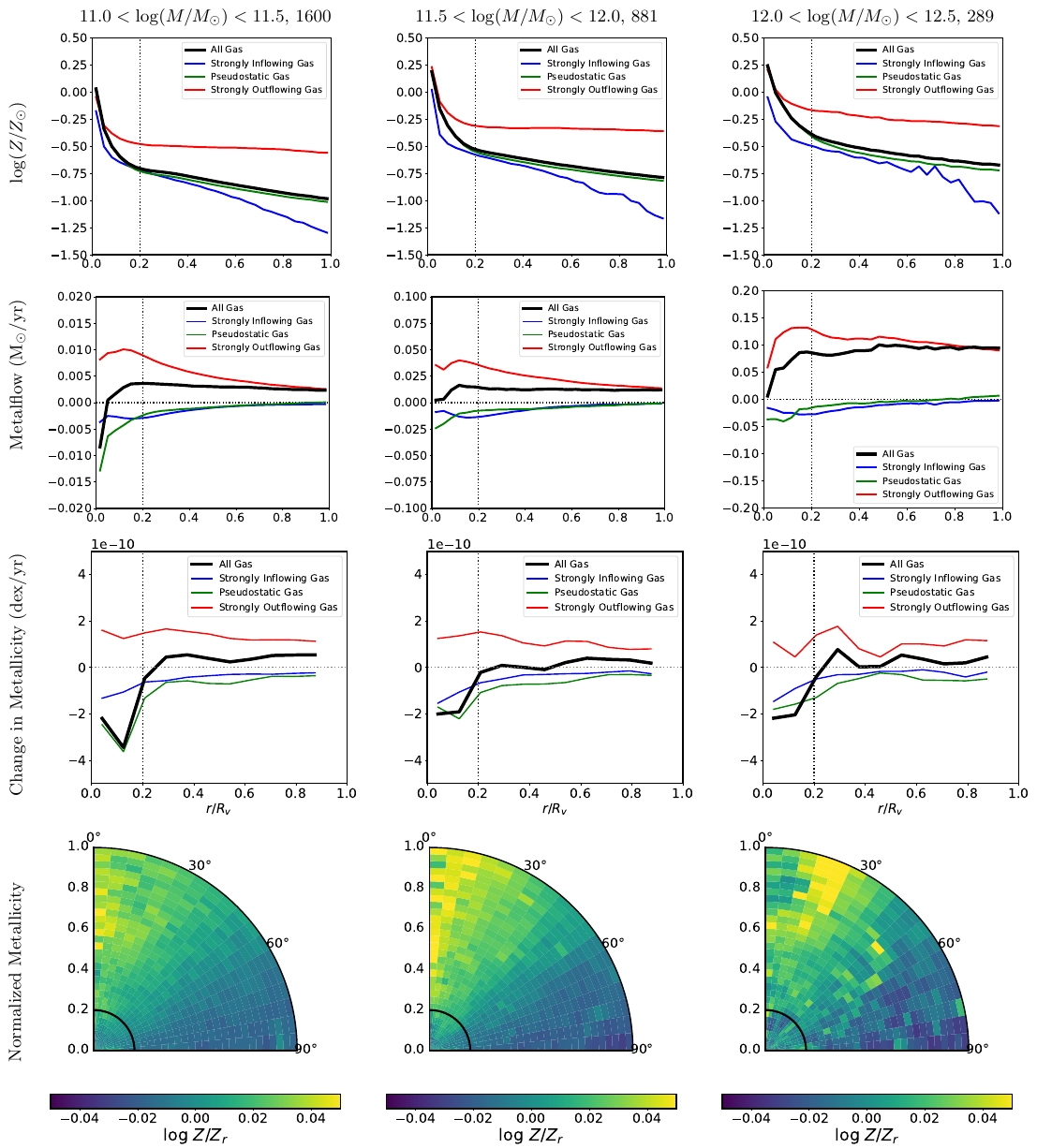}
\caption{Row 1: Radial profiles of metallicity for each gas type. Row 2: Metalflow. Row 3: Change in metallicity. Row 4: 2D metallicity normalized by radial profile.}
\label{fig:all_metals}
\end{figure*}

Figure \ref{fig:all_metals} shows the metallicity radial profiles, metalflow, and metallicity accumulation rates for our full sample. The metallicity profiles (row 1) are qualitatively the same across mass; a steep slope within $0.2R_v$, followed by a shallow slope. At all radii, the total gas metallicity closely matches the pseudostatic gas, and strongly inflowing material is slightly lower metallicity, while the strongly outflowing gas is markedly higher. The gas at low radius is well mixed (similar metallicity between strongly inflowing, pseudostatic, and strongly outflowing gasses) while the gas types are distinct at higher radius. The metallicity gradients past $0.2R_v$ are steepest for strongly inflowing gas, followed by pseudostatic gas, and strongly outflowing gas has a flat profile. The differences in pseudostatic metallicity profiles across galaxy type decreases with mass. In general, metallicity at all radii increases with mass for each feedback type (but see Section \ref{subsec:results_RM}). Row 2 of the figure shows the metalflow rates. There are many similarities between the mass and metalflows; in the first two mass brackets the pseudostatic and strongly inflowing gases transport roughly the same amount of metals inward and the flow rate for all types of gas decreases with radius; in our highest mass bracket the rate of transport by strongly outflowing gas is nearly constant with radius, resulting in large net outflows. Unlike the massflow, however, the net metalflow is outward at nearly all radii at all masses; the combination of high velocity and high metallicity means the strongly outflowing gas moves metals much more effectively, even though the average metallicity closely matches the pseudostatic gas.

Row 3 of the Figure shows the rate of change in metallicity. The large negative values seen at low radii in all three plots indicate that gas-phase metals are leaving the area, but does not account for the additional metal produced by stars. Given the negative metallicity profiles in our sample, local star formation must replenish these areas. Like the metallicity profiles, the change in the metallicity is a similar value at the same scaled radius for all gas types and all halo masses. Within $0.2R_v$, metals are being diluted; outside of $0.2R_v$, metals are accumulating. This is one of many transitions occur at the characteristic radius of $0.2R_v$, demarcating the interior region. In this interior region the pseudostatic gas dominates, and is responsible for a sharp decline in metallicity as pristine gas builds up in the center of the halo. Past this radius the influence of pseudostatic gas decreases and strongly outflowing gas is allowed to accumulate metals at a moderate rate. For all galaxy types at all radii, metallicity is reduced by strongly inflowing and pseudostatic gas, and replenished by outflowing gas. Without stars as a source of new metals, the inner metallicity would decrease by an average of $\sim 30 \%$ in the next Gyr. 

The fourth row of the Figure shows 2D maps of metallicity normalized by the radial average, $Z_r$. The black line marks $0.2R_v$. Several past works have found that the CGM has higher metallicity along the galaxy minor axis, ie, in the direction of outflow \citep{metangle, metnorm2D, minor_major}. We also detect a slight increase in CGM metallicity near the poles. Although this difference is sometimes visible within $0.2R_v$, as for our middle mass bracket, most halos have the same metallicity profile along both axes until higher radius. When splitting gas by dynamic type (not shown), we see that strongly outflowing and pseudostatic gas are higher metallicity in the polar regions, but strongly inflowing gas metallicity does not vary with angle. This is because a large amount of strongly inflowing gas is accreted from outside the halo.

\begin{figure*}
\includegraphics[width=1.0\textwidth]{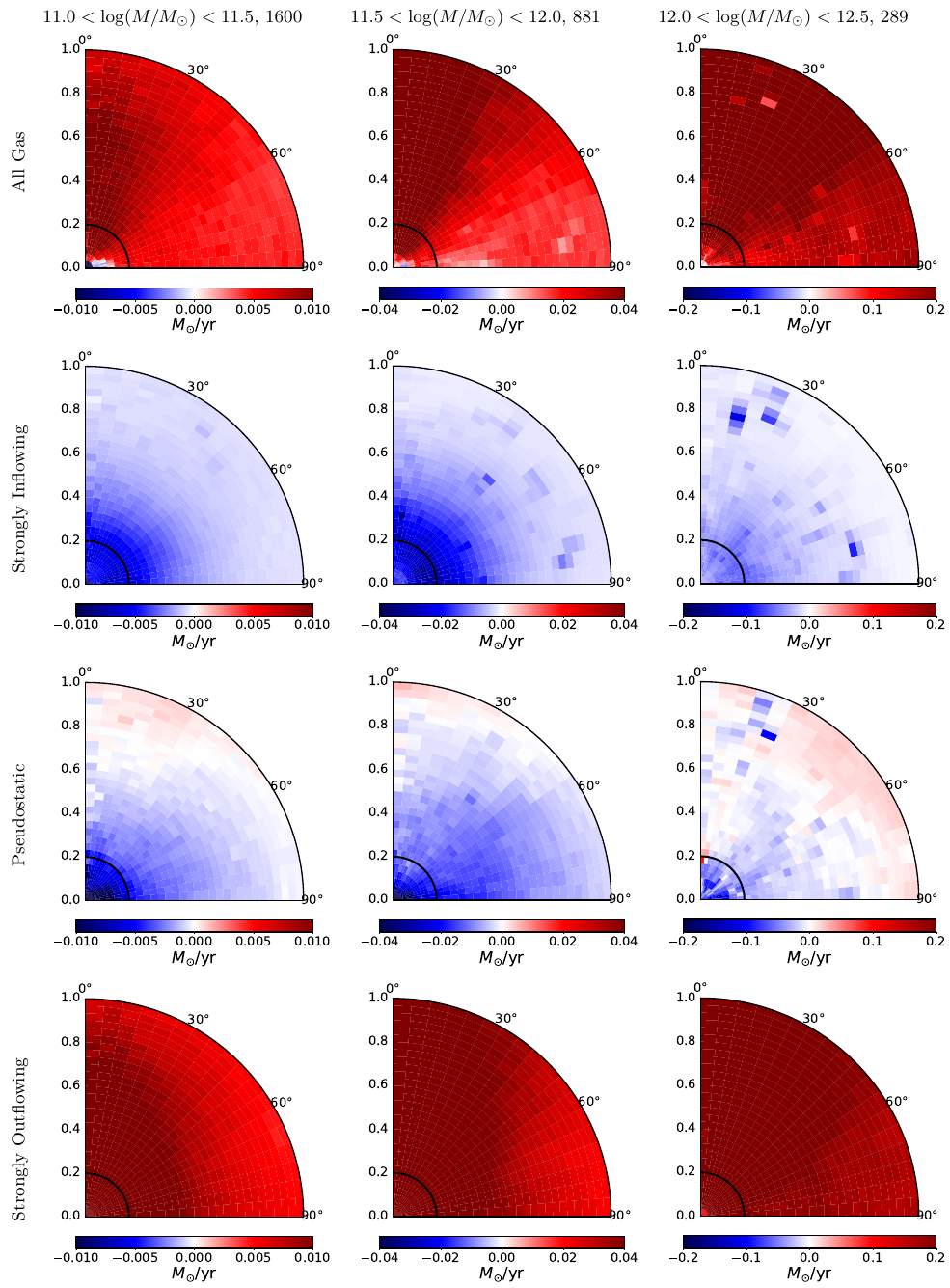}
\caption{As in Figure \ref{fig:all_2d_massflow}, but for the flow of gaseous metal instead of mass.}
\label{fig:all_2d_metalflow}
\end{figure*}

Figure \ref{fig:all_2d_metalflow} shows the 2D metalflow of our sample. When compared to Figure \ref{fig:all_2d_massflow}, we see that the strongly outflowing gas is much more important to the metalflow, and halos across our mass range have a net outflow of metal at all angles. Although the majority of metal is contained in pseudostatic gas, the metallic dynamics are mostly determined by the outflowing gas. The slowly accreting pseudostatic gas can create regions of neutral to light flow away from the poles, as seen in the middle mass bracket. Nearly all pseudostatic metals are inflowing. A flattened shape which may correspond to a disk is visible in the top row for the lower two mass brackets. However, the possible disk is not clear in the plots isolating inflowing, pseudostatic, or outflowing metal alone. Metals can inflow strongly within $0.2R_v$ of the galactic center in the lower two mass brackets, but this central region, and seemingly the ``disk'', is disrupted by the strong feedback in the most massive halos. 

\subsection{Results: The Effects of Star Formation} \label{subsec:results_SF}
\begin{figure*}[h]
\includegraphics[width=1.0\textwidth]{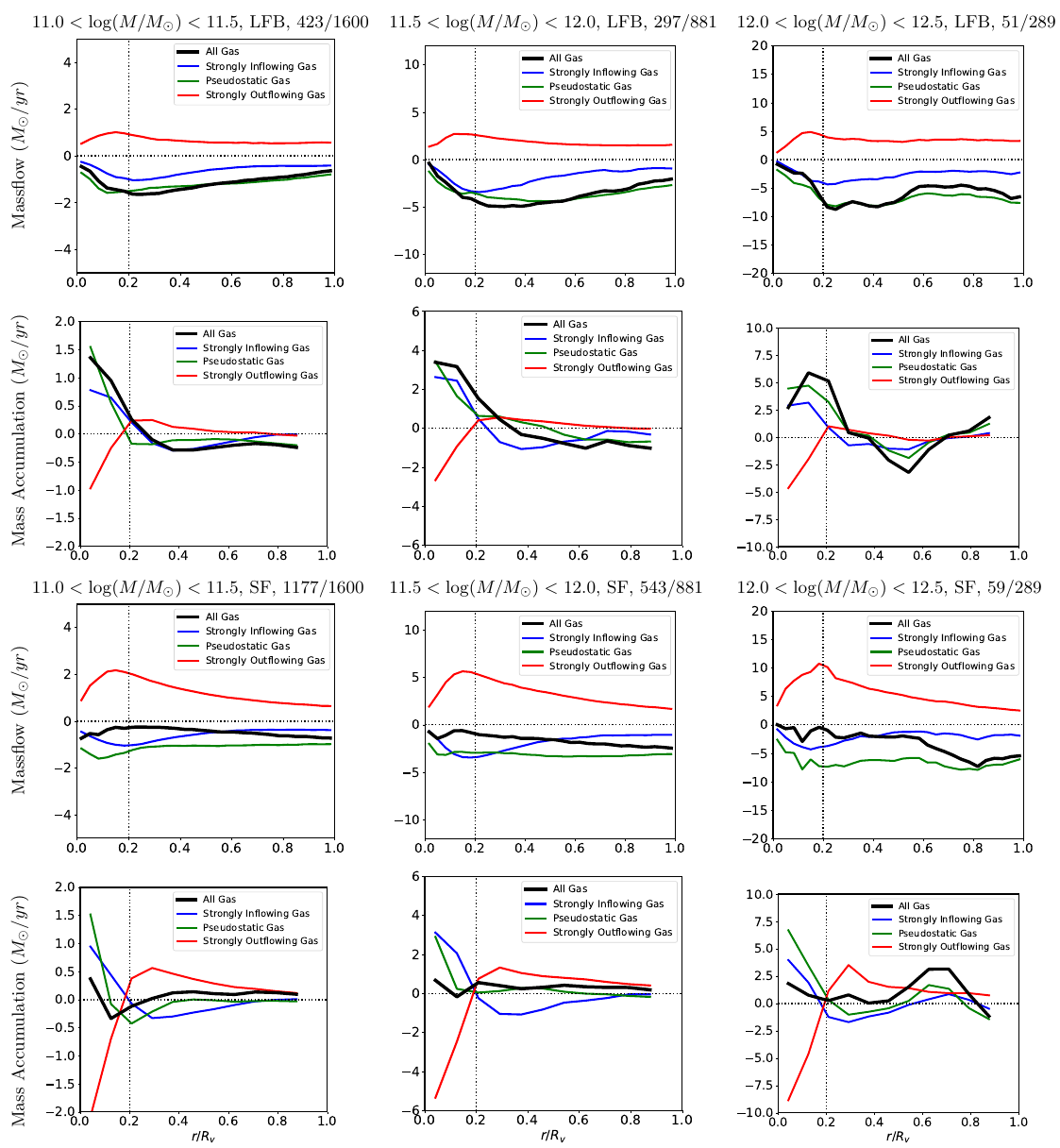} 
\caption{Massflow and mass accumulation of LFB halos (top two rows) and SF halos (bottom two rows). The numbers in the top right of each column heading show the fraction of halos in each mass range which have that feedback classification.}
\label{fig:SF_massflow_mdot}
\end{figure*}

In order to determine the effects of star formation feedback on gas flows, we compare halos with low feedback (LFB) to those that are actively star-forming (SF). LFB halos do not lack feedback entirely, they simply do not meet our star-forming or AGN activity thresholds. Figure \ref{fig:SF_massflow_mdot} shows the massflow (rows 1 and 3) and mass accumulation rate (rows 2 and 4) for LFB and SF halos. We see that the shape of the LFB and SF profiles are qualitatively different, while the shape remains similar across a wide range of halo masses for halos with the same feedback type. This pattern holds for RM and SFRM feedback as well (see Sections \ref{subsec:results_RM} and \ref{subsec:results_SFRM} respectively). Stacked halos with the same feedback type but different mass are significantly more similar than halos with different feedback types at the same mass.

In Figure \ref{fig:SF_massflow_mdot}, we see strongly outflowing gas is clearly influencing the total gas behavior of the SF halos, while the total behavior of the gas in LFB halos closely follows the pseudostatic gas. Looking at rows 1 and 3 of the Figure, the peak in the rate of mass transported by strongly flowing gas (in or out) occurs near $0.2R_v$. Quiescent and star-forming halos have a net inflow at all radii at all masses, and in each mass bracket, a similar amount of strongly inflowing and inflowing pseudostatic gas at large radii. However, for SF galaxies, strong outflows originating at the center of the halo prevent the rapid inflows seen in LFB halos, and cause the total gas flow to decouple from the pseudostatic flow. As a result the net inflow within $0.4R_v$ is much less in SF halos. 

In rows 2 and 4 of Figure \ref{fig:SF_massflow_mdot}, we can see the rise and decline of massflow from strongly flowing gas is caused by a pile-up of mass near the characteristic radius $0.2R_v$, where some of the gas seems to transition to pseudostatic, giving a flatter flow profile for this gas type. Star-forming halos are accumulating high metallicity, outflowing mass everywhere past this radius, leading to a total gas build up near $0.2$ and $0.6R_v$ for the highest mass halos. Most halos accumulate mass in their inner regions-- in the full sample this region is $\sim 0.2R_v$, but LFB halos have an expanded inner accumulation zone closer to $0.3-0.4R_v$ in size. The amplitude of the accumulation is also larger than for other galaxy types throughout this expanded region. Past $0.4R_v$ there is little to no accumulation, implying gas falls easily into the inner region for LFB halos. Although LFB halos quickly accumulate mass here, their radial mass profile is not significantly more concentrated than other galaxy types (see Appendix A), and there is little to no star formation. This suggests the strongly inflowing period is short-lived, and possibly a precursor state to star formation. The gas in the galactic center is emptied by high velocity outflows and replenished by inflowing and pseudostatic gas, determining the overall accumulation and size of the inner region. The star-forming sample shows a much more prominent accumulation of outflowing mass at $0.2R_v$, and a much more rapid emptying of the central regions by outflowing gas. This results in the innermost region of overall mass accumulation shrinking to $<0.1R_v$. When normalized by virial radius, we see the size of the central accumulating region, the shape of the massflow profiles, and CGM dynamics in general are determined primarily by feedback mode, not halo mass. 

\begin{figure*}[h]
\includegraphics[width=1.0\textwidth]{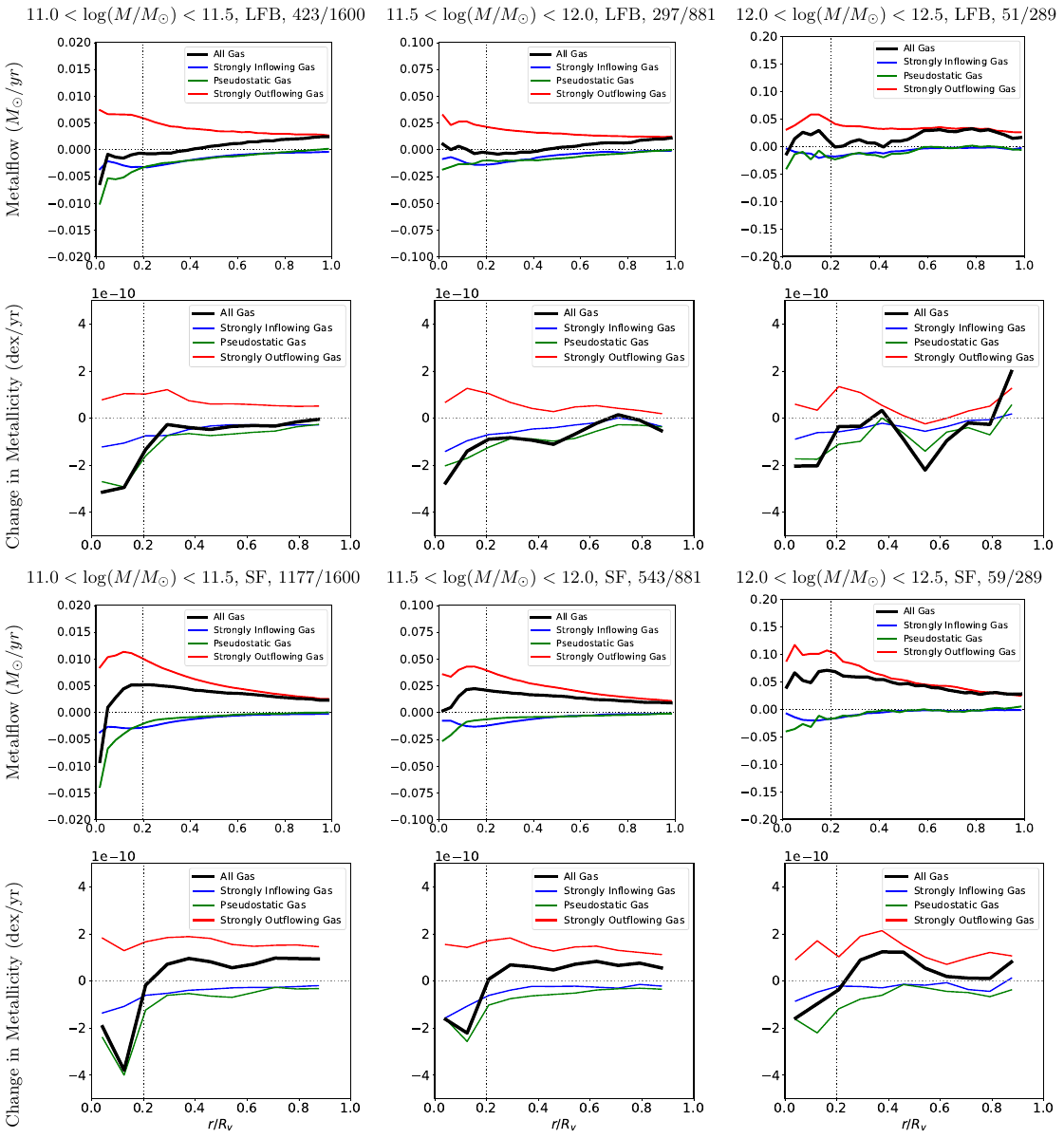}
\caption{As in Figure \ref{fig:SF_massflow_mdot}, but for metals.}
\label{fig:SF_metalflow_zfraclogdot}
\end{figure*}

\begin{figure*}[h]
\includegraphics[width=1.0\textwidth]{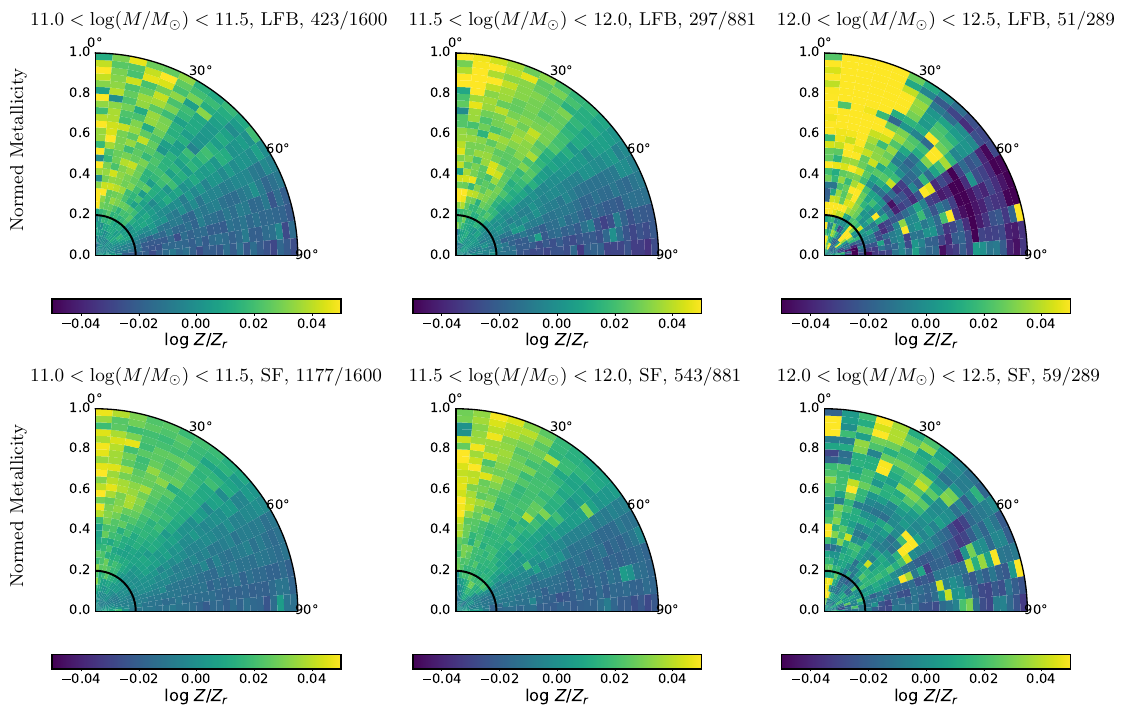}
\caption{2D metallicity, normalized by radial profile for LFB halos (top) and SF halos (bottom).}
\label{fig:SF_2D_normed_metallicity}
\end{figure*}

Figure \ref{fig:SF_metalflow_zfraclogdot} shows the metalflow (rows 1 and 3) and metallicity accumulation rate (rows 2 and 4) for LFB and SF galaxies. Again, there are qualitative similarities between the mass and metal dynamics; in both cases the flow rate for SF and LFB galaxies is similar at high radius, but at low radius the SF feedback causes strong outflows for metals and weak inflows for mass. Metals are preferentially outflowing compared to mass, but there are some areas in LFB galaxies in the lower mass brackets where there is so much inflowing gas that it causes a net inflow of metal, despite accreted gas being more pristine. Looking at rows 2 and 4, we see LFB halos are experiencing metal depletion at most radii, while SF feedback pushes metals out of the central region, where it accumulates out to the virial radius. Note that LFB halos are generally not expelling metals, but rather allowing for the unopposed accretion of low metallicity gas. Figure \ref{fig:SF_2D_normed_metallicity} shows the normalized metallicity maps for the SF and LFB sample. Although there are small differences, we do not find a correlation between feedback mode and angular metallicity structure. 

\begin{figure*}[h]
\includegraphics[width=1.0\textwidth]{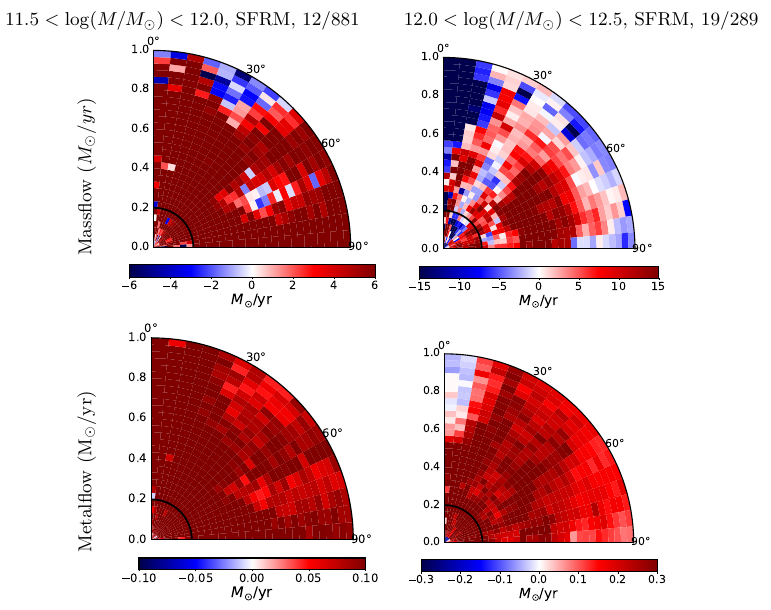}
\caption{2D massflow and metalflow of LFB halos (top two rows) and SF halos (bottom two rows).}
\label{fig:SF_2d_massflow_metalflow}
\end{figure*}

The two dimensional flow of mass for LFB and SF halos is shown in rows 1 and 3 of Figure \ref{fig:SF_2d_massflow_metalflow}. LFB halos show intense inflows everywhere, but less so at the poles. Meanwhile, star-forming halos in the lower two mass brackets show prominent polar outflows, and between the poles, a disk-plane can be seen with mostly pseudostatic, inflowing mass. The outflowing poles are less defined for the most massive SF halos. When comparing to the top row of Figure \ref{fig:all_2d_massflow}, we see the SF halos in the lower two mass bins bridge the gap between the LFB and general results. However, the activity seen in the most massive star-forming halos cannot explain the uniform and overwhelming outflow shown for our full sample in this mass range. When breaking down the massflow into inflowing, outflowing, and pesudostatic gases (not shown), we see a pattern very similar to our results in Figure \ref{fig:all_2d_massflow}, with a similar amount of strongly inflowing gas but less inflowing pseudostatic gas along the polar regions for SF galaxies.

The metallicity radial profile for our low feedback and star-forming samples are similar both to each other and to our full sample average, but with star-forming halos having very slightly lower metallicity in all gas types at all radii (Appendix A). However, there are significant differences in the two dimensional flow of metal. In the first row of Figure \ref{fig:SF_2d_massflow_metalflow}, LFB halos have a consistent inflow of mass, dominated by pseudostatic gas. There is slightly less inflow in the polar regions; in the rare cases of outflows without significant star-formation or AGN activity, the outflows are polar. Likewise, as seen in Figure \ref{fig:SF_massflow_mdot}, our SF sample is generally experiencing an inflow of mass, with strong outflows confined to the polar regions.

\clearpage

The inflowing and pseudostatic gasses have nearly the same effect on metallicity for our LFB and SF halos; they always cause a decrease in metallicity, and this decrease becomes more extreme within $0.2R_v$. However, the LFB halos are declining in metallicity at all radii, while metallicity is increasing past $0.2R_v$ for SF halos. The outflowing gas is the main differentiator, as in SF halos it causes larger increases in metallicity at all radii, leading to a net increase in metallicity in the outer regions. Note that the rate of change in metallicity is affected by the metallicity of gas already present-- pristine infalling gas mixing with pristine gas results in little change, whereas the same pristine infall mixing with metal-rich gas produces a large effect. So the outflowing gas in SF halos causes an increase in metallicity, even though it is less enriched than the outflowing gas of LFB halos (Appendix A). 

The metalflow of our LFB halos is notably different from the full sample. This is the only subsample in which we see a significant inflow of metal, which only occurs off-polar and is being funneled to the disk-plane. Turning to the bottom row of Figure \ref{fig:SF_2d_massflow_metalflow}, we see star-forming halos have the lowest metalflow in these off-polar regions. They also all have a disk-like area of neutral metal flow. Although this disk-like region is not visible for the global highest mass bin, the metalflow for the SF and full sample for the lower two mass bins are similar; nearly universal outflow with tiny core of inflowing metal and possible disk. This appears to be an of inversion of what we see for the flow of mass, which is inflows originating at high radius only interupted by outflowing mass from the inner galaxy. However, both are a result of central star-forming activity creating metals which can only move to even smaller radii if they inflow, and of pristine gas originating from outside the halo, so that mass is always generally inflowing and metal generally outflowing. The only exception to this rule are the high mass halos, which overall have outflows at large radii (Figure \ref{fig:all_massfrac_massflow_mdot}). The most massive star-forming halos actually have less outflowing metal and mass than the full sample for this mass range. This is because their behavior is dominated by strong AGN activity, which we discuss in the following section.




\clearpage

\subsection{Results: The Effects of Radio Mode Feedback} \label{subsec:results_RM}
In this section we analyze the mass and metalflow of halos experiencing significant radio mode (RM) feedback. As discussed in Section \ref{sec:TNG}, AGN feedback is tied closely with halo mass, and there is only one RM halo in the lowest mass bracket. RM AGN are also rare in the middle mass bracket, with only 29/881 meeting our feedback threshold. However, radio-mode halos are the most common type in the highest mass bracket with 160/289 meeting our threshold. In this section, we present the results for the middle and upper mass bracket only, and regard the middle bracket as a qualitative check for the most massive halos.

\begin{figure*}[h]
\includegraphics[width=1.0\textwidth]{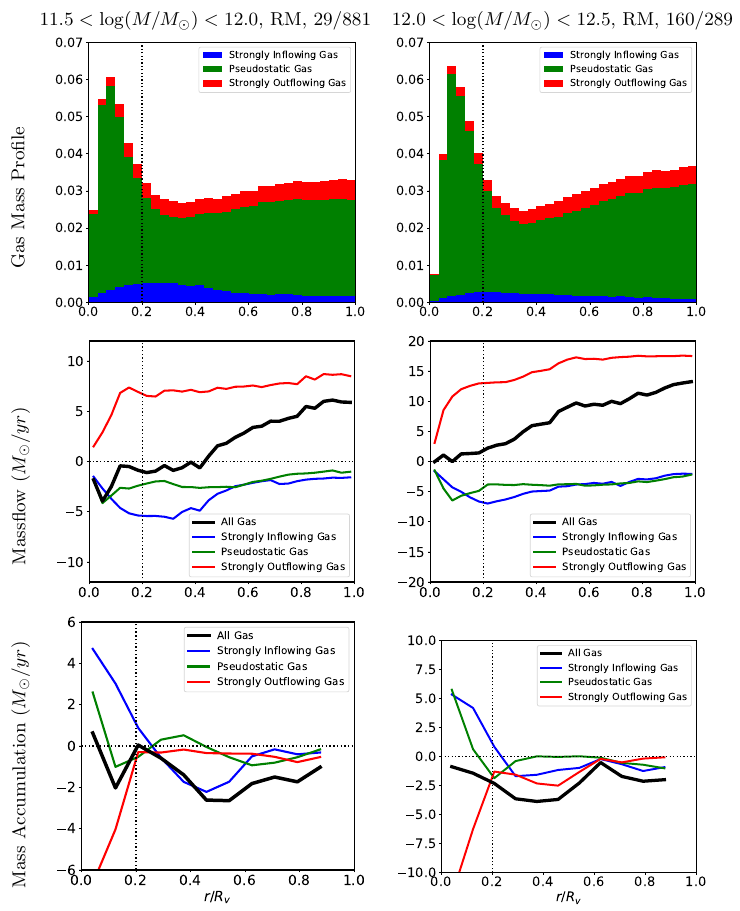}
\caption{Radial distribution of gas mass (top), flow of mass (middle), and mass accumulation rate (bottom) for RM halos.}
\label{fig:RM_massfrac_massflow_mdot}
\end{figure*}

In Figure \ref{fig:RM_massfrac_massflow_mdot} we see that the mid-mass and high mass RM sample have many similarities. In the top row, the RM halos have an enhanced strongly outflowing gas profile throughout when compared to the top row of Figure \ref{fig:all_massfrac_massflow_mdot}, a ``knee'' in the gas PDF between $0.2-0.4R_v$, and a positive slope past this radius-- similar also to the mass PDF of the most massive halos in our full sample. In the middle row of the figure, the flow rates of strongly inflowing gas and inflowing pseudostatic gas are equal at high radius, while at lower radii strongly inflowing gas dominates, especially at mid-mass. However, the most dominant dynamic gas type is the strongly outflowing gas. The total massflow of the LFB and SF sample closely tracked the pseudostatic gas, and featured a net inflow at all radii. The massflow rate of all gas types in LFB and SF halos decreases with radius or is roughly constant past $0.2R_v$ (see Figure \ref{fig:SF_massflow_mdot}). However, in RM halos, the flow rate of strongly outflowing gas reaches a high value at $0.2R_v$ and continues to slowly increase with radius, causing this subsample to have a large outflow past $0.4R_v$ in both mass ranges, and no net inflow whatsoever for the highest mass halos. The slight inflow within $0.4R_v$ for the mid-mass halos is caused by a comparable strong inflow rate at smaller radii that is not present in the highest mass bracket. In the bottom row we see RM halos have a much larger accumulation of strongly inflowing gas within $\sim 0.2R_v$ than SF or LFB galaxies and a similar accumulation rate for pseudostatic gas. However, there is no net central mass accumulation due to the extreme outflows caused by the AGN. In general, mass is being lost at all radii. 

The radial profiles of high mass RM halos look similar to the results from the full sample (Figure \ref{fig:all_massfrac_massflow_mdot}). This is because over $55 \%$ of the high mass sample are RM halos, while very few are in the lower two mass brackets; the apparent differences with mass seen in Section \ref{subsec:results_all} are due to RM feedback becoming more important at high mass.  

\begin{figure*}[h]
\includegraphics[width=1.0\textwidth]{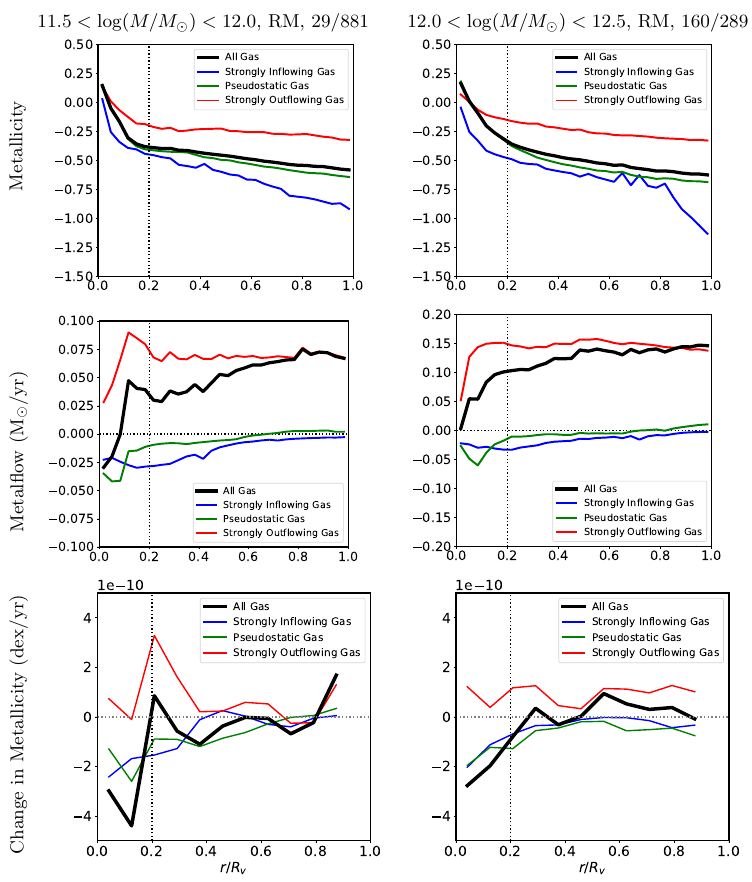}
\caption{Metallicity (top), metalflow (middle) and metallicity accumulation (bottom) for RM halos.}
\label{fig:RM_metallicity_metalflow_zfraclogdot}
\end{figure*}

The metallicity profiles, metalflow, and metallicity accumulation rate for RM halos are shown in Figure \ref{fig:RM_metallicity_metalflow_zfraclogdot}. As discussed in Section \ref{subsec:results_all}, the metallicity profiles across our sample are quite similar. Although there are differences between LFB and SF halos, the main differentiator is RM feedback. Metallicity profiles essentially come in two types; RM-quiet halos have a large concentration of metal in the innermost bin and a steeper slope within $0.2R_v$, while RM-loud halos have a lower peak metallicity and a flatter slope within this region. This results in a total metallicity drop of $\sim 0.4$dex from the innermost point to the virial radius, while other galaxy types have a drop of $\sim 1$ regardless of mass range. When compared to other halo types, the strongly outflowing gas in RM halos is the least enriched and the pseudostatic and inflowing gasses are the most enriched, suggesting the metal mixes into the outer CGM despite kinetic differences. Looking at row 2 of the Figure, both the high velocity outflowing gas and total gas behavior reach their peaks in metal transport at a lower radii than their peaks in mass transport, and have a flatter profile that does not continue to increase at high radius. This is because metals are preferentially located in the inner regions. When comparing the bottom rows of Figure \ref{fig:SF_metalflow_zfraclogdot} and Figure \ref{fig:RM_metallicity_metalflow_zfraclogdot}, we see SF feedback in the highest mass bracket causes a peak in the rate of change of metallicity within $0.6R_v$, whereas RM halos are losing metallicity at this radius, and have a smaller peak in metallicity accumulation outside this radius. In LFB halos the decrease in central metallicity is due to the unopposed infall of pristine gas, while in RM halos it is caused by AGN feedback pushing metals to higher radius, even past the virial radius.

\begin{figure*}[h]
\includegraphics[width=1.0\textwidth]{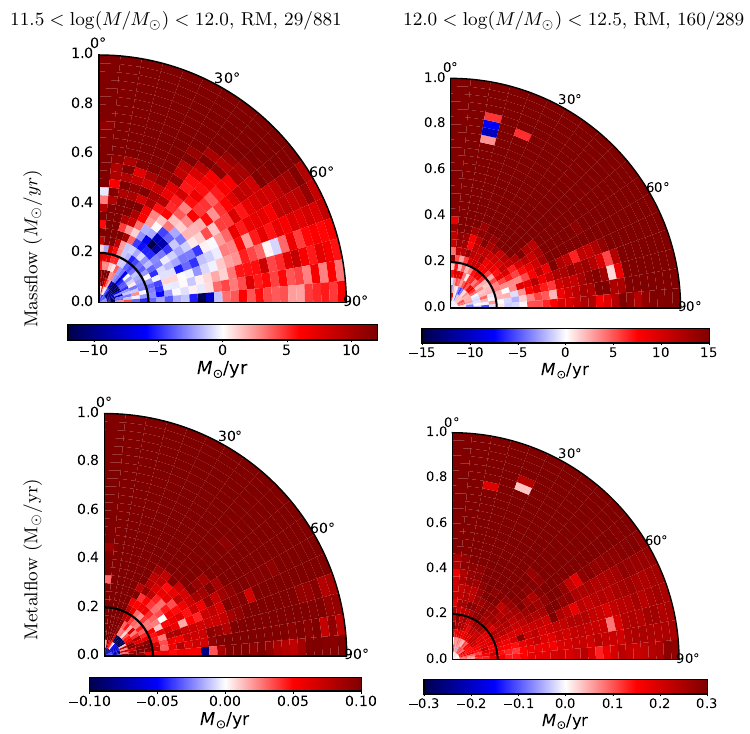}
\caption{2D massflow (top) and metalflow (bottom) for RM halos. Note the difference in colorbar normalization compared to previous sections (.1 vs .04 $\mathrm{M_{\odot}/yr}$ at mid-mass, .3 vs .2 $\mathrm{M_{\odot}/yr}$ at high mass).}
\label{fig:RM_2d_massflow_metalflow}
\end{figure*}

Figure \ref{fig:RM_2d_massflow_metalflow} shows the 2D massflow and metalflow of the radio mode sample. Mass is outflowing at all angles, but not all radii; within $0.4R_v$ there is no net flow, or even a net inflow for the mid-mass bracket, and the outflows intensify with radius, reflecting what we see in Figure \ref{fig:RM_massfrac_massflow_mdot}. The central inflowing/neutral mass regions and metal regions are not cospatial by coincidence, but because the metalflow is generally following the pseudostatic gas. This behavior is more obvious in the SFRM halos, where we have an off-center mass/metal inflow (see Figure \ref{fig:SFRM_2d_massflow_metalflow}).

\clearpage

\subsection{Results: The Effects of Star Formation and Radio Mode Feedback} 
\label{subsec:results_SFRM}
We now turn our attention to halos with both star formation and AGN feedback.  These halos are an extreme minority compared to our other classifications, with 0/12/20 in the lowest/middle/highest mass brackets, but their behavior is informative and a check on what we discovered about feedback response in previous sections. We note this is not necessarily the same as the halos with the ``most'' feedback. For instance, strong kinetic feedback quenches star formation, and SFRM halos tend to have less RM feedback than quenched RM halos. 

\begin{figure*}[h]
\includegraphics[width=1.0\textwidth]{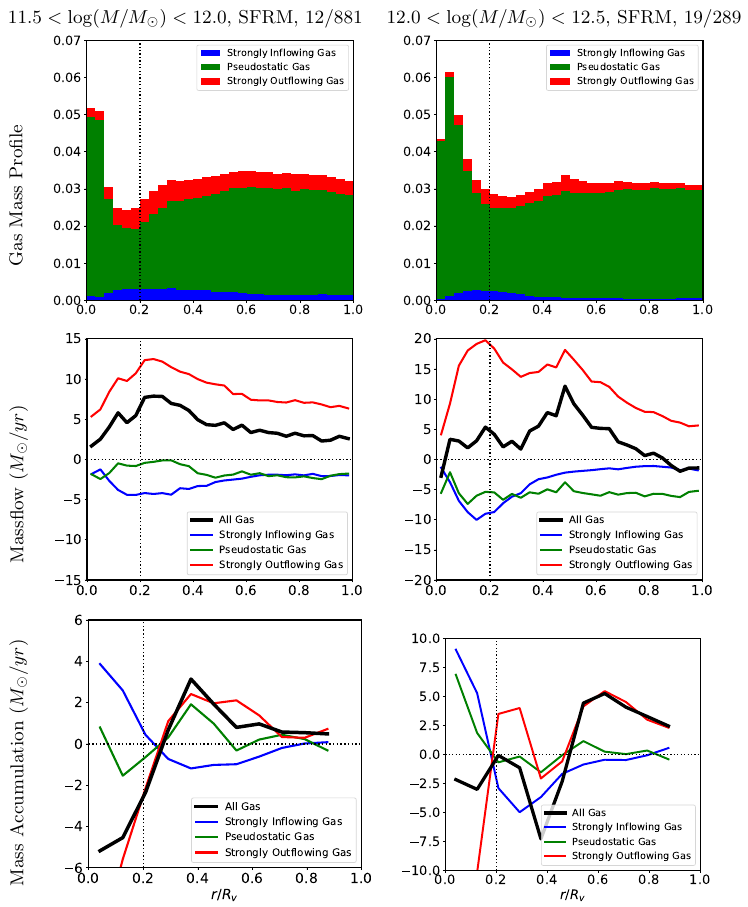}
\caption{Radial distribution of gas mass (top), flow of mass (middle), and mass accumulation rate (bottom) for SFRM halos. Notice the increased y-limits for the massflow of the middle mass-bracket compared to previous sections (15 vs 10 $\mathrm{M_{\odot}/yr}$).}
\label{fig:SFRM_massfrac_massflow_mdot}
\end{figure*}

Figure \ref{fig:SFRM_massfrac_massflow_mdot} shows the gas mass profile, massflow, and mass accumulation radial profiles for this sample. In the top row, we can already see familiar features from Figure \ref{fig:RM_massfrac_massflow_mdot}. The SFRM halos, especially for the middle mass bracket, have a large amount of outflowing gas and a pronounced knee at $0.2R_v$, similar to RM halos. However, the highest mass halos have much less strongly outflowing gas and a flat total gas profile past $0.4R_v$, more similar to the profile of star-forming halos (Appendix A). The mid-mass halos have some strongly inflowing gas throughout, as they do for RM halos, but high mass halos have very little inflowing gas. 

When compared to the RM halos, the massflow plots show the amount of gas transported by pseudostatic and inflowing gas is similar at all radii for the mid-mass halos, but the higher mass sample has significantly more inflowing pseudostatic gas past $0.4R_v$ and more strongly inflowing gas within $0.4R_v$. The central net inflowing region within $0.4R_v$ for the mid-mass RM halos is gone, and instead this is the region of peak net outflow. Within this region is also the bump in outflowing gas at $\sim 0.2R_v$ which is much more pronounced for all star-forming halos. We find that SF and SFRM halos have a denser CGM, which may be preventing the outflowing gas from escaping the inner CGM. The radial profiles of the outflowing gas look like superpositions of the outflowing profiles for SF and RM halos. Overall, the strongly outflowing gas transports a huge amount of mass and dominates the motion of the flow, as in kinetic mode halos. There is also a strong net outflow at most radii, though the amplitude is less. However, both the strong outflow and net outflow decline with radius past $0.6R_v$ ($0.4R_v$ for mid-mass halos), with a peak near $0.2R_v$, similar to the SF sample. The highest mass halos even have a slight inflow past $0.8R_v$. 

Because of the change in slope of the massflow at $0.2R_v$ (and between $0.4-0.6R_v$ for the high mass sample), there is a pile-up of outflowing mass at this radius, seen in the bottom row of the Figure. As with radio mode halos, the inner region is evacuated by strong outflows and partially replenished by strong inflows and pseudostatic gas. However, RM halos do not accumulate mass at almost any radii, and appear to be shedding it overall. The situation is less clear for SFRM halos; the mid-mass halos have an accumulation of mass near $0.4R_v$ that is as high or higher than the total central mass accumulation in the rest of our sample in this mass range. But mass is both inflowing and outflowing from this radius in the high mass sample, rapidly depleting it. Instead there is a strong accumulation closer to $0.6R_v$, which is also as high or higher than the total central mass accumulation for all other halos in this bracket. A key difference between RM and SFRM halos is this large accumulation of mass at high radii. 

\begin{figure*}[h]
\includegraphics[width=1.0\textwidth]{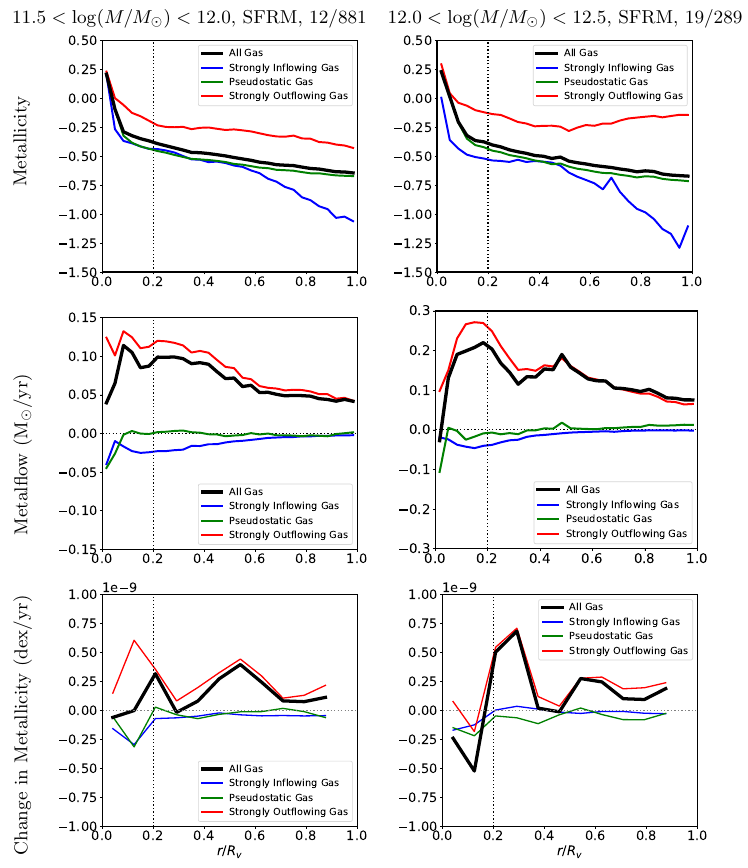}
\caption{Metallicity (top), metalflow (middle) and metallicity accumulation (bottom) for SFRM halos. Note the increased y-limits of the metallicity accumulation compared to previous sections ($10^{-9}$ vs $5 \times 10^{-10}$).}
\label{fig:SFRM_metallicity_metalflow_zfraclogdot}
\end{figure*}

Figure \ref{fig:SFRM_metallicity_metalflow_zfraclogdot} shows the metallicity radial profile, metalflow, and metallicity accumulation for SFRM halos. LFB, SF, (Appendix A) and SFRM halos all have a steep inner metallicity gradient indicating a dense inner region. Although all halo types transition from a steep gradient to a more shallow one at $0.2R_v$, the outer gradient is  steeper for SF and SFRM halos compared to LFB and RM halos. In the case of LFB halos this is because the lack of feedback allows gas to collect quickly in the center, while for RM halos this is because the black hole rarifies the central gas and ejects metals to higher radii. This ejection still occurs for SFRM halos, but a denser inner region must be maintained for star formation to occur. Although it is not enough to change the total metallicity profile (black line), the outflowing gas of SFRM halos carries so much metal that the metallicity profile has a positive slope past $0.6R_v$. This is the only type of gas in the only type of halo to exhibit a positive metallicity gradient.

The rate of change in metallicity is significantly higher for SFRM halos than the other feedback categories due to a combination of metal expulsion (primarily from star formation) and gas rarefication (primarily from RM feedback). When comparing the metalflow and metallicity accumulation rates to the massflow and mass accumulation rates of SFRM halos (row 2 of Figures \ref{fig:SFRM_massfrac_massflow_mdot} and \ref{fig:SFRM_metallicity_metalflow_zfraclogdot}) we see the shape of the profiles is roughly the same; dominated by extreme outflows that remain strong to high radii, with strongly inflowing gas transporting more material than pseudostatic gas at lower radius. However, the total metalflow tracks the strongly outflowing material more closely than in the massflow plots. Looking at the metallicity accumulation rates (row 3 of Figure \ref{fig:SFRM_metallicity_metalflow_zfraclogdot}), we see both mass brackets have a large peak near $0.2R_v$ and $0.6R_v$. These correspond to the radii where outflowing mass is accumulating. In particular, compare the large increases in metallicity near $0.2R_v$ and $0.6R_v$ in the highest mass bin. In the mass accumulation plots, the peak near $0.6R_v$ is larger, and the peak near $0.2R_v$ exists only for strongly outflowing gas, while other gas types are leaving the area, especially the strongly inflowing gas. However, in the metallicity accumulation plots, the peak at $0.2R_v$ is the larger. This is because at $0.6R_v$ the mass of lower metallicity inflowing and pseudostatic gas is not changing by much, whereas at $0.2R_v$ there is both an accumulation of high metallicity gas and a decrease in low metallicity gas.

\begin{figure*}[h]
\includegraphics[width=1.0\textwidth]{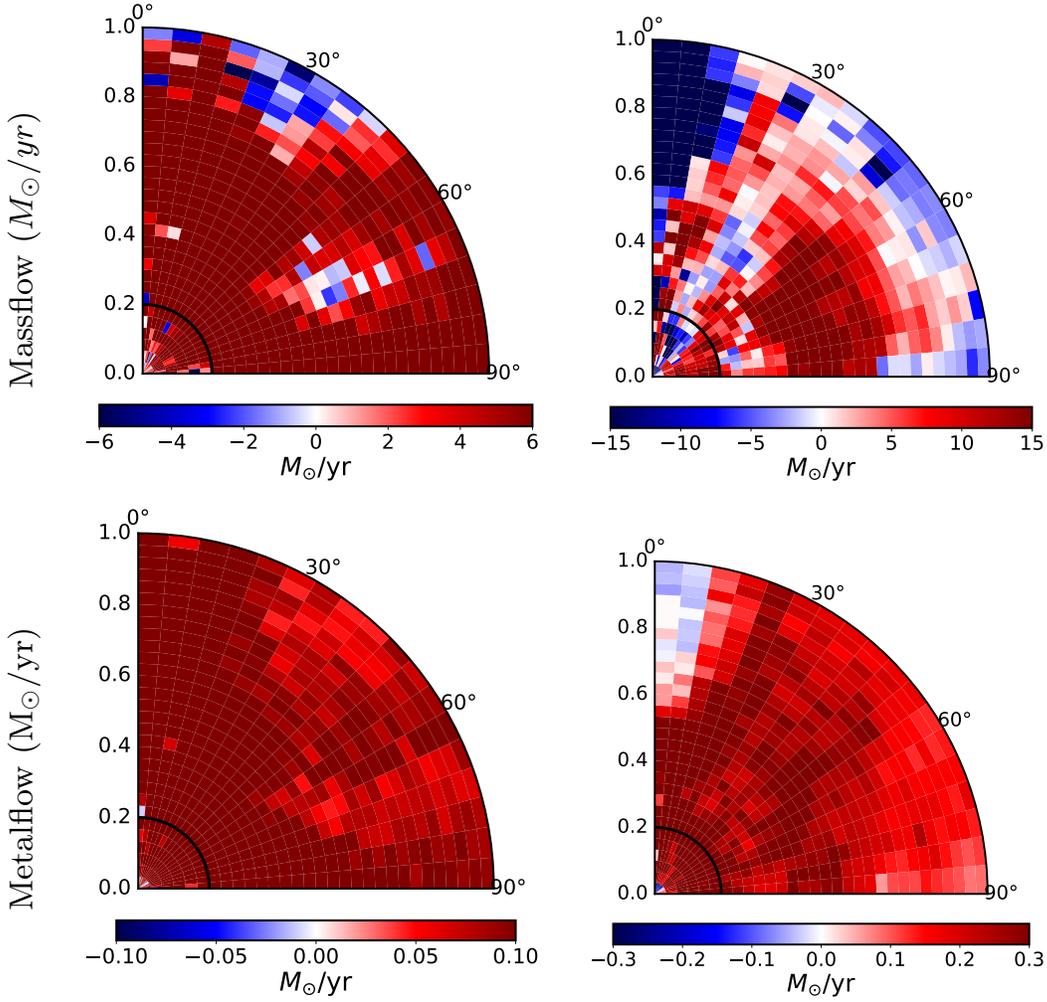}
\caption{2D massflow (top) and metalflow (bottom) for SFRM halos. Note the difference in colorbar normalization compared to previous sections (.1 vs .04 $\mathrm{M_{\odot}/yr}$ at mid-mass, .3 vs .2 $\mathrm{M_{\odot}/yr}$ at high mass).}
\label{fig:SFRM_2d_massflow_metalflow}
\end{figure*}

Figure \ref{fig:SFRM_2d_massflow_metalflow} shows the 2D massflow and metalflow for SFRM halos. These halos have significant inflowing regions near the poles, with the high mass sample having a strong inflow that penetrates from the virial radius to the central region. This is caused by a combination of small sample size and minor mergers. These mergers do not preferentially happen at the poles, but they do preferentially happen to SFRM halos, implying some of these halos have star formation triggered by a satellite. Aside from minor mergers, in the highest mass range these halos also have a nearly uniform slight inflow at high radius, even though on average halos at this mass do not have any inflow. From the top rows in Figure \ref{fig:SFRM_2d_massflow_metalflow}, Figure \ref{fig:SF_2d_massflow_metalflow}, and \ref{fig:RM_2d_massflow_metalflow}, we conclude that a strong net inflow of gas at high radius is necessary but not sufficient for star formation. At the bottom of the Figure we see that metals are almost always outflowing, as the gas accreted from high radii is more pristine. However there is a metallic inflow near the pole at high mass which corresponds to a satellite. 

\subsection{Results: The Effects of Morphology}
\label{subsec:morphology}
\begin{figure*}[h]
\includegraphics[width=1.0\textwidth]{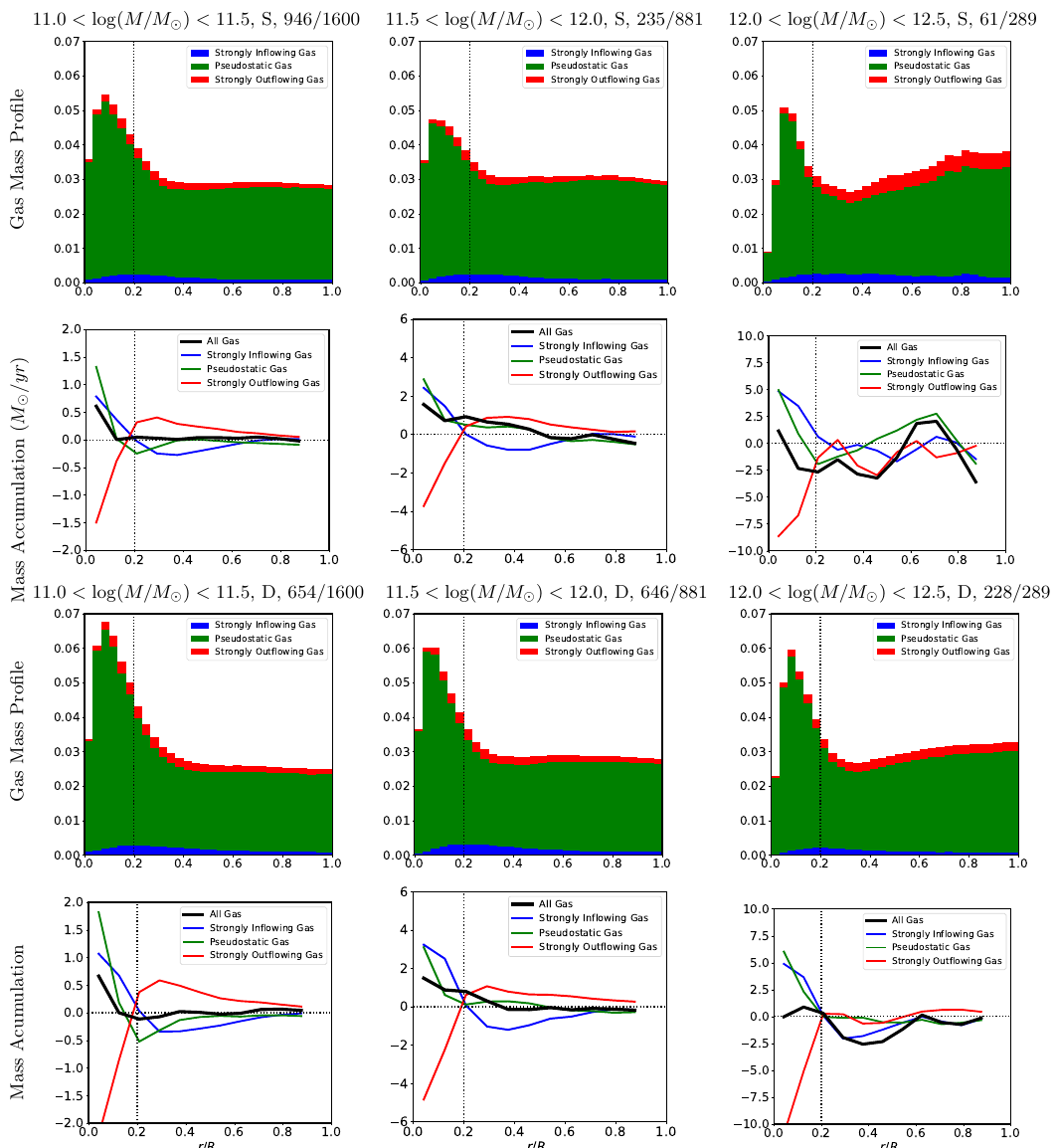}
\caption{Gas mass profile and mass accumulation rate for spheroidal halos (top two rows) and disks (bottom two rows).}
\label{fig:morph_massfrac_mdot}
\end{figure*}

In this section we explore the morphology of our sample. The radial mass distribution and mass accumulation of disks (D) and spheroids (S) are shown in Figure \ref{fig:morph_massfrac_mdot}. In rows 1 and 3 of the Figure, we see that the gas mass of disk galaxies is more centrally concentrated than that of spheroids, but past $0.2R_v$ their profiles are quite similar, with low and mid-mass halos having flat distributions and high mass halos having a rising distribution past $0.3R_v$. At high mass, spheroidal halos have more strongly flowing gas throughout and a more sharply rising profile at high radius. Rows 2 and 4 show that most halos have little net accumulation outside the central region except for in the middle mass bracket, and particularly for spheroidal halos. High mass spheroids also have an apparent accumulation of pseudostatic material between $0.6R_v$ and $0.8R_v$ due to the motion of a satellite. Both spheroids and disks are losing significant mass between $0.2-0.6R_v$, but only high mass spheroids are losing mass within $0.2R_v$, where high mass disks are neutral or accumulating material. 

The radial massflow, metallicity, metalflow, and metallicity accumulation profiles (Appendix A) are similar to each other, and to our results in section \ref{subsec:results_all}. Between $11.0<\log(M/M_{\odot})<12.0$, the radial flow of material is qualitatively the same, though disk galaxies tend to have slightly higher metallicity and more extreme values. At high mass, spheroids have a slightly rising tail. These differences decrease with halo mass and with radius, where the morphology of the central galaxy is less important.  

\begin{figure*}[h]
\includegraphics[width=1.0\textwidth]{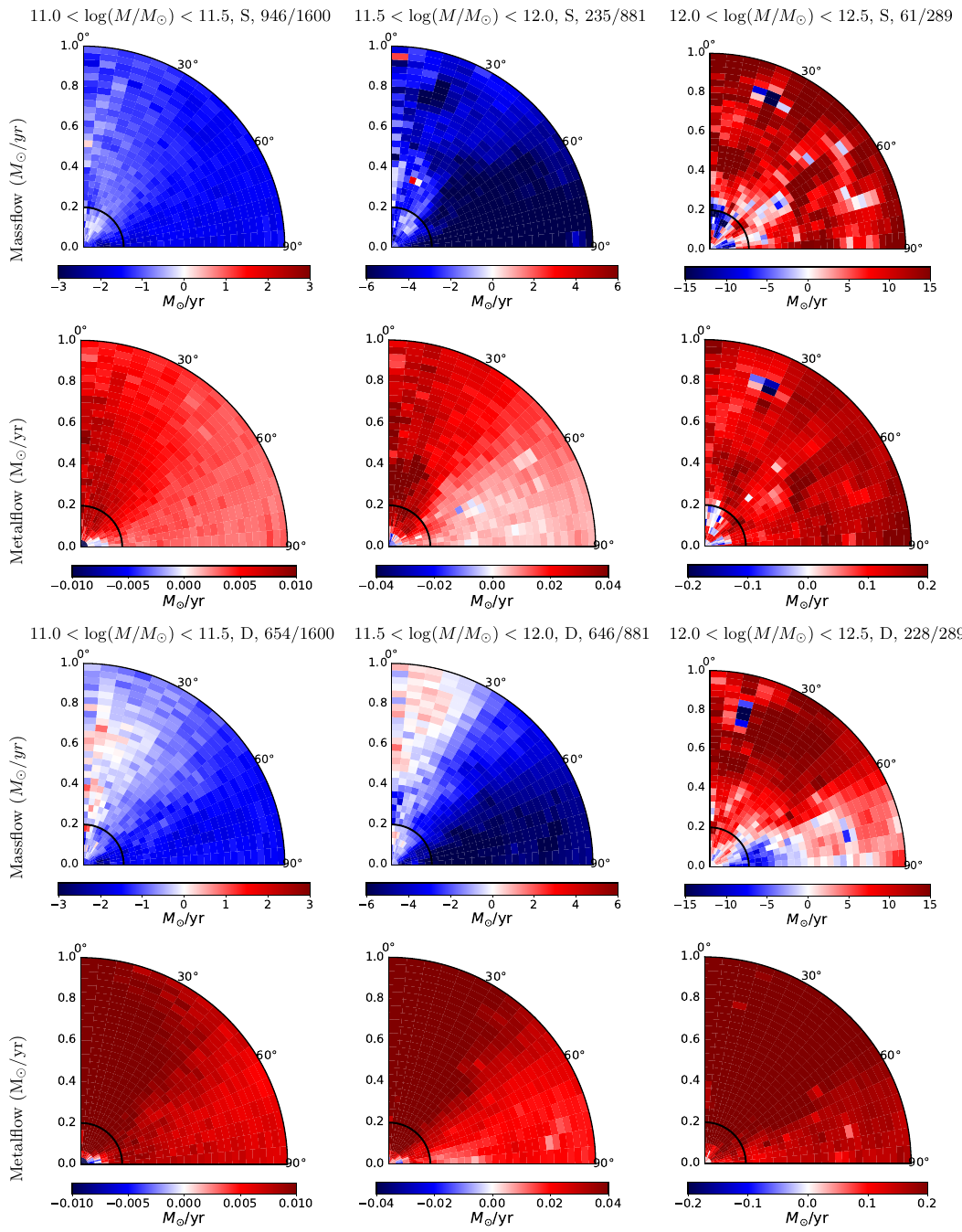}
\caption{2D massflow and metalflow for spheroids (top two rows) and disks (bottom two rows).}
\label{fig:morph_massflow_metalflow}
\end{figure*}

Although the radial profiles of disks and spheroids are often similar, disks have more angular structure at all masses. The 2D mass and metalflow of our sample is shown in Figure \ref{fig:morph_massflow_metalflow}. In general, the flow of mass and metal is  higher in disks than spheroids. The massflow in the lower two mass brackets is significantly more bipolar in disks, with neutral and outflows confined to distinct cones that reach out to the virial radius. However, the percent of SF disks and spheroids is similar in these mass brackets, with $76/72 \%$ and $63/60 \%$ of disks/spheroids are classified as SF for the lowest and middle mass range, respectively. The distribution of star formation rate of disks and spheroids is also similar (Appendix A). This implies disk galaxies respond differently to feedback than spheroids. Low to mid-mass disks and spheroids have similar radial profiles, but morphology can effect angular flow distribution out to the virial radius. 

The highest mass halos have large outflows everywhere, disrupted by pockets of neutral and inflows. For disk galaxies, these pockets are prefferentially aligned to the disk, confined to ``reverse cones'', also reaching the virial radius. The metalflow plots show bipolar structure in both disks and spheroids in the lower two mass brackets, but only disk galaxies have faint but constrained ``reverse cones'' of less intensely outflowing metal in the highest mass bracket. There are small disks of neutral and inflowing metal with radius $<0.2R_v$ visible in low mass and mid-mass disk galaxies, as well as in low mass spheroids.

\section{Discussion}
\label{sec:discussion}

\begin{figure*}[h]
\hspace{-0.5cm}
\includegraphics[width=1.0\textwidth]{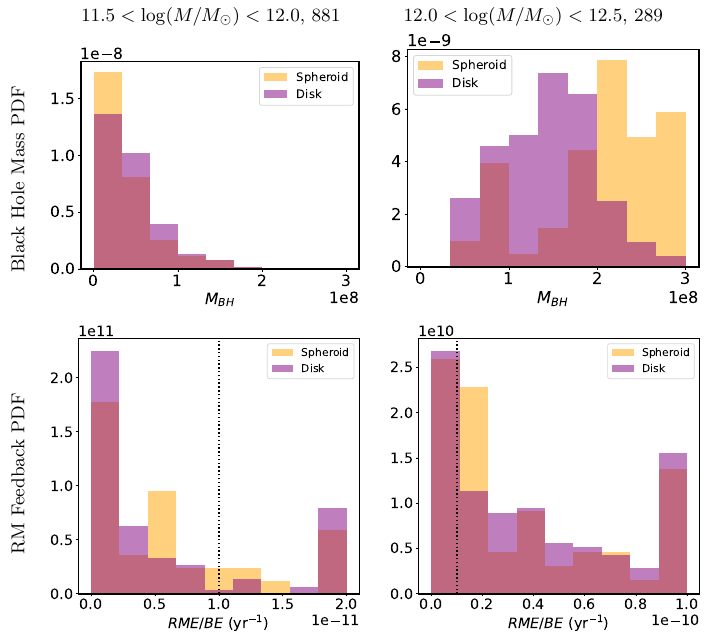}
\caption{Histograms of black hole mass (top) and RM activity (bottom) for disks and spheroids. In the bottom row, the vertical dotted line marks our RM activity threshold, and a long tail in the high mass distribution of disk galaxies has been truncated and all placed in the highest bin. The results when divided by feedback type are similar-- exceptions discussed in text.}
\label{fig:BHmass_RM_morphology}
\end{figure*}

In terms of feedback and gas flow, it is useful to divide our sample into two primary subgroups; RM and non-RM, and then to compare the effects of star formation on these two  groups. From Figures \ref{fig:SF_massflow_mdot} and \ref{fig:RM_massfrac_massflow_mdot} we see that SF (SFRM) halos have less inflow (more outflow) and less mass accumulation within $0.2R_v$ due to the feedback. However, SF and SFRM halos must get fuel from somewhere, and we see they both have net inflows near the virial radius, even in the presence of powerful RM feedback. Similarly, SF (SFRM) halos have an enhanced metallicity gradient within $0.2R_v$ compared to LFB (RM) halos. In general, the inner region is dominated by SF feedback, and the outer region dominated by RM feedback, save the caveat that net inflow near $R_v$ is necessary for star formation. This is consistent with \citet{fcgm}, who found that a more massive CGM correlates with star formation rate. Although this was not a good discriminator for our sample, we found that CGM density was, and that SF (SFRM) halos have a denser CGM than LFB (RM) halos, with RM having the least dense CGM. 

In this study we find that differences in gas dynamics are not directly linked to halo mass, but that different mass regimes are dominated by different feedback modes. Halos falling into the same feedback mode have extremely similar mass and metal flows regardless of halo mass. 

In the lowest mass bracket there is no radio mode feedback, and in the second only $4\%$ of halos have any RM activity. We essentially have LFB and SF halos only, and in similar proportions ($26/74 \%$ and $34/62 \%$ LFB/SF for the lowest and middle mass brackets, respectively), thus the gas dynamics in these mass ranges is quite similar. But in the highest mass range $55\%$ of halos have RM feedback, which causes large differences in the average behavior of the full sample. The massflow in the lowest and middle mass brackets is largely determined by the pseudostatic gas, with the strongly inflowing/outflowing gas mostly responsible only for the bipolar shape. But the highest mass halos are governed by the powerful outflows from RM activity. 

Both RM feedback and SF feedback can slow the infall of gas, but only RM feedback can entirely quench a halo. Plots of the massflow (see section \ref{sec:results}) show key differences between the quenching RM outflows and the SF outflows. star-forming halos actually have net mass accumulation beyond $0.4R_v$, and the highest mass SF halos have their peak in mass accumulation near $0.6R_v$. However, lest we mistake the pile-up of gas outside the characteristic radius as a general effect of feedback, we see the kinetic mode halos completely lack this, and in fact mass is leaving the region in these halos at all radii, except for within $<0.1R_v$ for the mid-mass range. This is an important difference between the star-forming and kinetic AGN feedback; although both can push gas out of the star-forming central regions, the weaker SF FB only pushes gas to within the central CGM, where the high metallicity gas may cool and form stars again, while AGN feedback is able to beam the mass and metals through the inner CGM, dispersing the gas and heating the outer regions to increase cooling times. Effectively, star formation feedback dominates the inner CGM, while radio mode feedback dominates the outer CGM. 

We also explored the effects of morphology. In general, there is little morphological dependence in the first two mass brackets. The fraction of disk galaxies changes from $41\%$ to for $11.0<\log(M_{halo}/M_{\odot})<11.5$ halos to  $73\%$ for $11.5<\log(M_{halo}/M_{\odot})<12.0$ halos, but the average gas flows are similar. Meanwhile, the highest mass brackets has a similar percentage of disks as the middle ($71\%$), but with notable angular differences between disks and spheroids. This switch in the importance of morphology coincides with the onset of RM activity; we posit that morphology is important in deciding how the galaxy reacts to RM feedback.

First we should establish that differences in morphology decide how a halo reacts to RM feedback, and not the other way around. \citet{morphology_sf_set} studied how morphology changed over time for galaxies with stellar mass $10^9-10^{11.5}M_{\odot}$ in the TNG simulation. They find that at all masses morphology is set during the star-forming and growth period of the galaxy, and not altered significantly by quenching. Our results are in agreement with this, as the disk fraction changes from total halo mass range $10^{11.0}-10^{11.5}$ to $10^{11.5}-10^{12.0}$, but not from $10^{11.5}-10^{12.0}$ to $10^{12.0}-10^{12.5}$, even though this is when RM-driven quenching begins.

The star-forming fraction of disks and spheroids is similar in the first two mass brackets, though disks are star-forming slightly more often. However, in the highest mass range, the SF fraction in general decreases and disks are more than twice as likely to be classified as SF than spheroids ($23 \%$ vs $11 \%$). Although the SFRM halos are split evenly between disks and spheroids in the intermediate mass bracket, they are overall a very small percent of total halos in this mass range ($\sim 1\%$). In the highest mass bracket they are more common, and all SFRM halos in this mass range are disks. Given this, SFRM halos no longer seem like strange outliers, but make up over $8\%$ of disk galaxies with $10^{12.0}M_{\odot}<M_{halo}<10^{12.5}M_{\odot}$, similar to the proportion of SF spheroids in this range. Combining SF and SFRM disks vs SF and (nonexistent) SFRM spheroids in the upper mass bracket, $32\%$ of disks are star-forming compared to $11\%$ of spheroids. Additionally, black hole-related quenching in TNG happens by rarifying the gas in the central region \citep{ejective_preventative}, and Figure \ref{fig:morph_massfrac_mdot} shows that high mass disks have a slight accumulation of mass within $0.2R_v$, while high mass spheroids are rapidly losing material in this region, save for in the very innermost bin. 

This suggests disks may have some resistance to quenching in TNG. \citet{quenchtime} studied over 12000 TNG subhalos with $M>10^9M_{\odot}$ and examined the relationship between pre-quenching conditions and the speed of the quenching process. They found quenching was slower in galaxies with larger: virial mass, central black hole mass, stellar halfmass radii, and specific angular momentum of accreted gas. We reserve an analysis of the angular momentum of our sample for another paper, but it has been found that the angular momentum of CGM gas and central halo are correlated \citep{AM}, and observations show that spiral galaxies retain a larger amount of their CGM angular momentum than ellipticals \citep{AM_spirals, AM_ellipticals}. Our results are in partial agreement with \citet{quenchtime}, as the quenching-resistant disks at all masses and of all feedback types have a larger stellar halfmass radius than the corresponding spheroidals. In our sample, for masses $10^{11.0}<M_{halo}/M_{\odot}<10^{12.0}$, disk galaxies of all feedback types are more massive. The situation is less clear for the highest mass halos, where LFB spheroidals are larger than LFB disks, but no strong correlation is found otherwise. We found no meaningful correlation between morphology and CGM specific angular momentum in our sample, and found that quenched halos had larger black holes, which is more in line with earlier results \citet{BH_quench}. However, we note that we are comparing the behavior of currently quenched RM-loud halos, and not their states pre-quenching.

Figure \ref{fig:BHmass_RM_morphology} shows histograms of black hole mass and radio activity for disks vs spheroids. Spheroids have more massive black holes than disks for most feedback types, which suggests spheroidal halos are more effective at growing black holes. Although there are only 7 SF spheroids in the highest mass bracket, they have significantly less massive black holes than SF disks, which suggests that disks can hold more massive black holes without quenching. When comparing black hole feedback levels directly, we see a similar pattern; SF spheroidals are cutoff in RM activity at approximately half our threshold value, while SF disks have a range of values extending to our cutoff threshold, implying the SFRM disks are an extension of this distribution. We note that differences in high mass SF halos may be caused by the low number of SF spheroids. However, when splitting only by morphology, the tail of the most extreme RM feedback consists entirely of disks, despite spheroids having more massive black holes. Then disk galaxies grow black holes more slowly, can host more massive black holes, can receive more feedback from those black holes before quenching, and can fuel more extreme radio mode feedback after quenching. At the same black hole mass in the same accretion mode the feedback increases linearly with accretion rate \citep{ejective_preventative}, so RM feedback is essentially a measure of the average black hole accretion rate over the last $200$~Myrs, meaning disks are accreting material more quickly. If disks are more effective at funneling mass toward their centers it could explain both the heightened RM activity at lower black hole mass and the more durable star-forming activity. This would imply spheroidals must have had higher accretion rates at some point over $200$~Myrs in the past, in order to have more massive black holes now.

\section{Conclusions}
\label{sec:conclusions}
In this paper we see that feedback is essential to understanding the flow of gas in galaxies of all types and a wide range of masses. We find that LFB (low feedback) halos have large amounts of inflowing gas from all angles, and even large regions of metalflow in the off-polar areas; SF (star-forming) halos have strong inflows originating at high radius and bipolar outflows of both mass and metal; RM halos have extreme outflows at all angles and lose mass and metal to regions past the virial radius, causing higher metallicity inflows. One of our most important results is that the CGM dynamics are much more sensitive to feedback mode than halo mass; differences which appear to be due to halo mass can be explained by SF feedback being more common at low and mid-mass and RM (radio mode) feedback being more common at high mass. 

We also examined a unique set of SFRM (star-forming and radio-loud) halos. The behavior of these halos is qualitatively a linear combination of SF and RM behavior. They feature net inflows at high radius, a steeper inner metallicity profile, and a peak in outflowing gas near $0.2R_v$ similar to SF halos. They also feature extremely powerful outflows that extend to high radius, before declining into a net inflow which allows for star formation. Similarly, the gas mass profile of the SFRM sample features a large amount of outflowing gas and prominent knee at $0.2R_v$ akin to the RM sample, but past $0.4R_v$ it does not increase with radius, but flattens, and the outflowing gas fraction declines to the same level as that found in SF halos (Appendix A).

We have reviewed several correlations between the gas and metal dynamics of the CGM and properties of the central galaxy. We have shown that the radius $0.2R_v$ is commonly a dividing line or inflection point for the mass profile, massflow, metalflow, mass and metallicity accumulation, and metallicity profiles for many galaxies, separating the ``inner behavior'' and ``outer behavior''. This radius may serve as a rule of thumb for dividing the central galaxy from the CGM. For the Milky Way, this corresponds to a characteristic radius of $\approx 40$~kpc. 

It is common to use metallic tracers such as magnesium to track the flow of material in halos. However, metallicity is determined by the motion and mixing of both metal-rich and pristine gas, which can result in surprising situations. In Section \ref{subsec:results_SFRM} we see that a similar accumulation of high metallicity outflowing gas has notably different effects on the overall metallicity in the region depending on the accumulation rate of pristine inflowing gas.

We showed in Section \ref{sec:results} that star formation feedback governs the inner CGM and radio mode feedback governs the outer CGM in TNG. Because inflowing gas is necessary for star formation, their relationship can be described as follows: the radio mode feedback determines if the outer CGM is inflowing; the flow state of the outer CGM determines if the galaxy is star-forming; star formation feedback determines the flow state of the inner CGM.

%






\begin{acknowledgments}
Acknowledgments: This material is based upon work supported by the National Science Foundation under Grant No. AST-2007013.
\end{acknowledgments}

\clearpage

\appendix

\section{Supplementary Plots}
\begin{figure*}[h]
\includegraphics[width=1.0\textwidth]{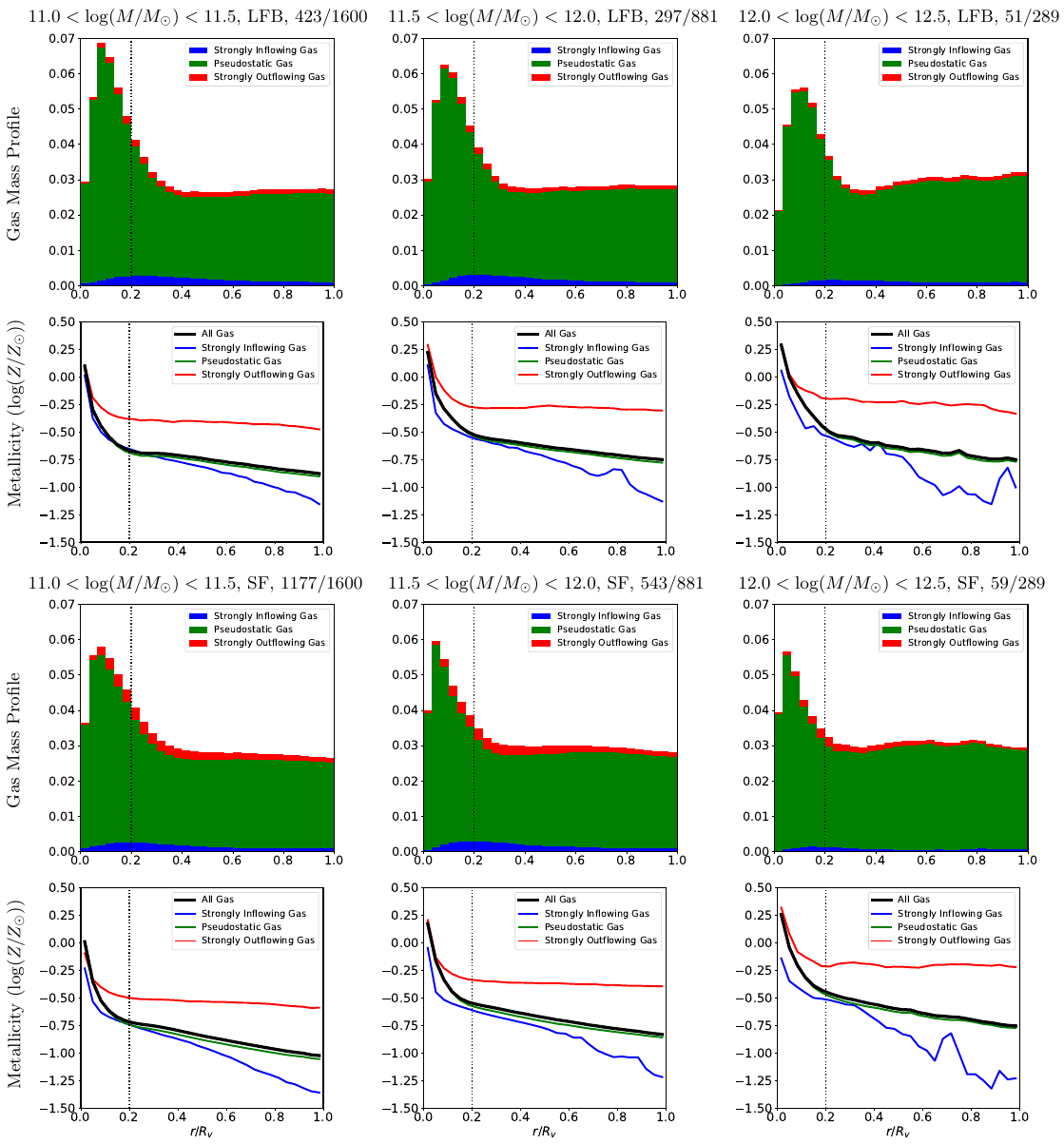}
\caption{Gas mass and metallicity profiles for LFB halos (top two rows) and SF halos (bottom two rows).}
\label{fig:SF_mass_metallicity_profile}
\end{figure*}

\begin{figure*}[h]
\includegraphics[width=1.0\textwidth]{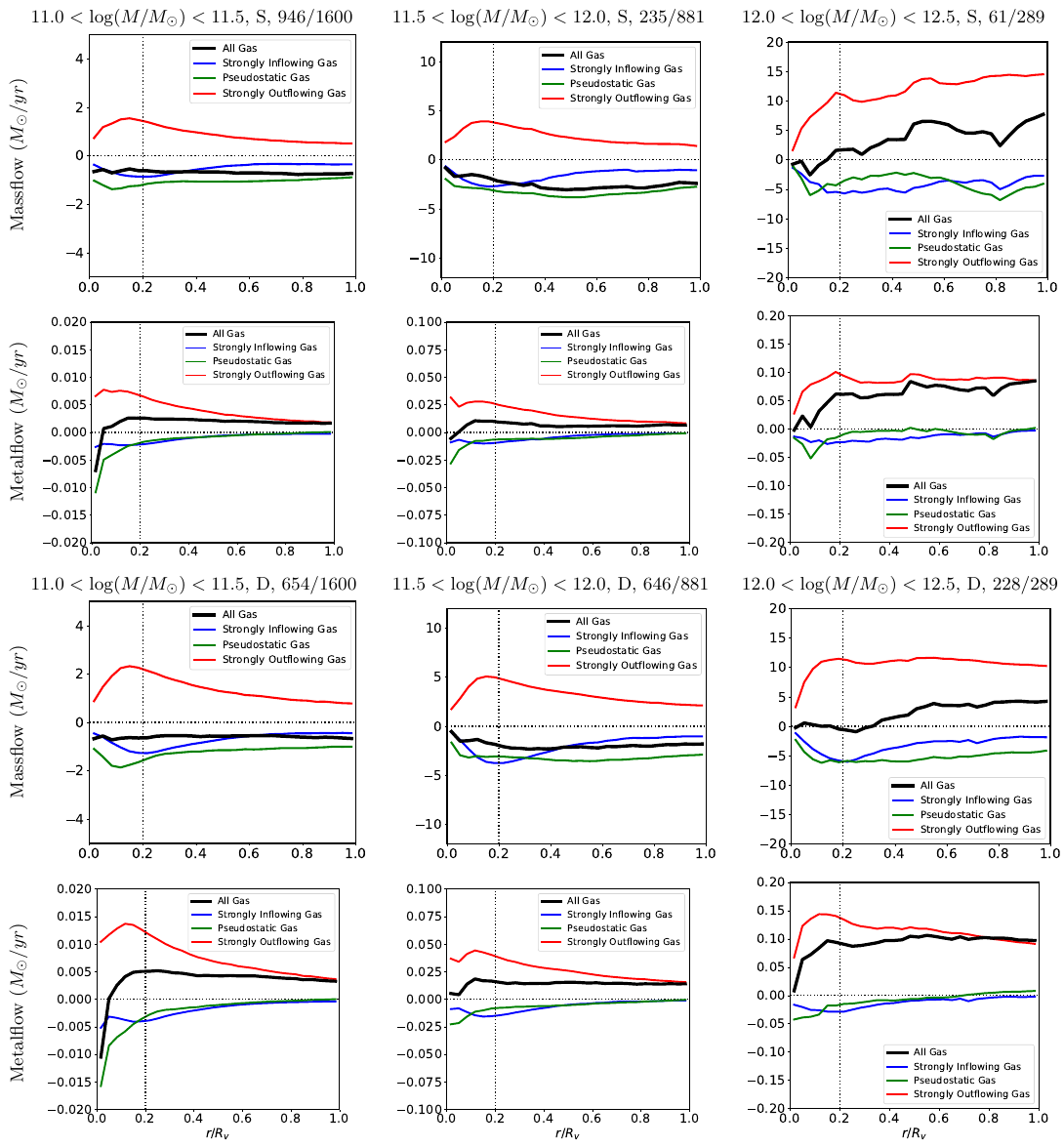}
\caption{Massflow and metalflow profiles for spheroids (top two rows) and disks (bottom two rows).}
\label{fig:Disk_massflow_metalflow}
\end{figure*}

\begin{figure*}[h]
\includegraphics[width=1.0\textwidth]{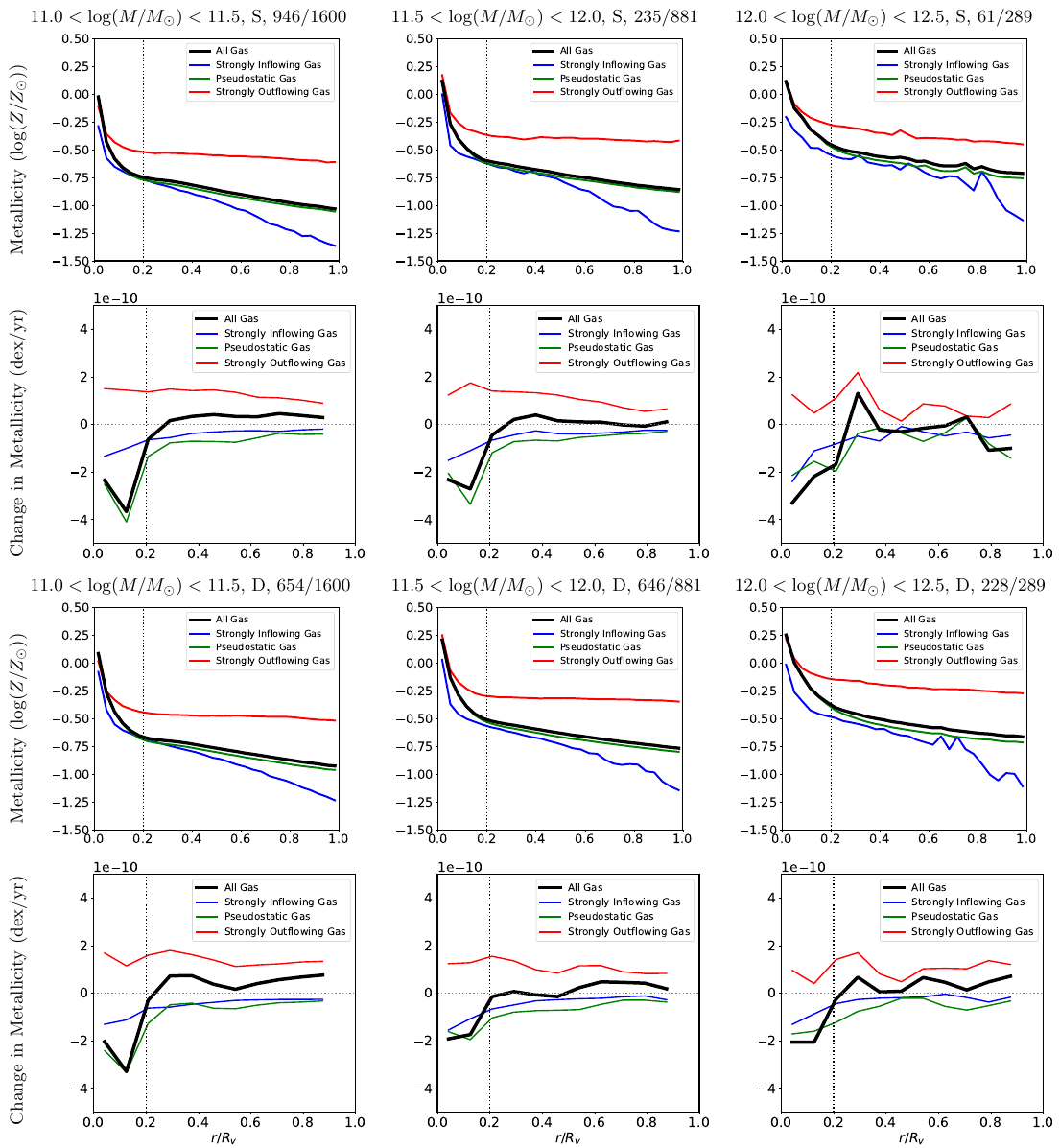}
\caption{Metallicity and metallicity accumulation profiles for spheroids (top two rows) and disks (bottom two rows).}
\label{fig:Disk_metallicity_zfraclogdot}
\end{figure*}

The gas mass profile and metallicity profile of SF and LFB halos is shown in Figure \ref{fig:SF_mass_metallicity_profile}. The radial massflow and metalflow of disks and spheroids is shown in Figure \ref{fig:Disk_massflow_metalflow}, and the radial metallicity and metallicity accumulation rates are shown in Figure \ref{fig:Disk_metallicity_zfraclogdot}.

\clearpage

\bibliography{Massflows}{}
\bibliographystyle{aasjournal}



\end{document}